\documentclass[fleqn,usenatbib]{mnras}

\usepackage[T1]{fontenc}

\DeclareRobustCommand{\VAN}[3]{#2}
\let\VANthebibliography\thebibliography
\def\thebibliography{\DeclareRobustCommand{\VAN}[3]{##3}\VANthebibliography}

\usepackage{amsmath,amstext,physics}
\usepackage[export]{adjustbox}
\usepackage{graphicx}
\usepackage{printlen,enumitem}
\usepackage{txfonts} 
\usepackage[figure,figure*]{hypcap} 
\usepackage{float}
\usepackage{sidecap}
\graphicspath{{./}{figures/}}

\newcommand{\nf}{{\bf\textsc{noAGN}}}
\newcommand{\mf}{{\bf m4e20}}

\newcommand{\simone}{{\bf m0.1e0.5}}
\newcommand{\simtwo}{{\bf m1e5}}
\newcommand{\simthree}{{\bf m2e10}}
\newcommand{\simfour}{{\bf m4e20}}
\newcommand{\simfive}{{\bf m10e50}}
\newcommand\msun[1]{{\rm M$_{\odot}$}}
\newcommand{\Msun}{{\rm M}_{\odot}}
\newcommand{\Msunh}{{\rm M}_{\odot}\,h^{-1}}
\newcommand{\Msunyr}{{\rm M}_{\odot}\,{\rm yr}^{-1}}
\newcommand{\Msunyrkpc}{{\rm M}_{\odot}\,{\rm yr}^{-1}\,{\rm kpc}^{-2}}
\newcommand{\Msunpc}{{\rm M}_{\odot}\,{\rm pc}^{-2}}
\newcommand{\kms}{{\rm km}\,{\rm s}^{-1}}

\newcommand{\sigmawind}{$\Sigma_{\rm wind}$}

\title[Dense stellar clumps from quasar winds]{Dense stellar clump formation driven by strong quasar winds in the FIRE cosmological hydrodynamic simulations}

\author[J. Mercedes-Feliz et al.]{Jonathan~Mercedes-Feliz,$^{1}$\thanks{E-mail: jonathan.mercedes\_feliz@uconn.edu}
Daniel~Angl{\'e}s-Alc{\'a}zar,$^{1,2}$
Boon~Kiat Oh,$^{1}$
Christopher~C. Hayward,$^{2}$\newauthor
Rachel~K. Cochrane,$^{3,2}$
Alexander J. Richings,$^{4,5}$
Claude-Andr{\'e} Faucher-Gigu{\`e}re,$^{6}$
Sarah Wellons,$^{7}$\newauthor
Bryan~A. Terrazas,$^{8}$
Jorge Moreno,$^{9}$
Kung Yi Su,$^{10,3,2}$
and Philip F. Hopkins$^{11}$
\\
$^{1}$Department of Physics, University of Connecticut, 196 Auditorium Road, U-3046, Storrs, CT 06269-3046, USA\\
$^{2}$Center for Computational Astrophysics, Flatiron Institute, 162 5th Avenue, New York NY 10010, USA\\
$^{3}$Department of Astronomy, Columbia University, 550 West 120th Street, New York, NY 10027, USA\\
$^{4}$E. A. Milne Centre for Astrophysics, Department of Physics and Mathematics, University of Hull, Cottingham Road, Hull, HU6 7RX, UK\\
$^{5}$DAIM, University of Hull, Cottingham Road, Hull, HU6 7RX, UK\\
$^{6}$CIERA and Department of Physics and Astronomy, Northwestern University, 1800 Sherman Ave., Evanston, IL 60201, USA\\
$^{7}$Department of Astronomy, Van Vleck Observatory, Wesleyan University, 96 Foss Hill Drive, Middletown, CT 06459, USA\\
$^{8}$Columbia Astrophysics Laboratory, Columbia University, 550 West 120th Street, New York, NY 10027, USA\\
$^{9}$Department of Physics and Astronomy, Pomona College, 333 N. College Way, Claremont, CA 91711, USA\\
$^{10}$Black Hole Initiative, Harvard University, 20 Garden St., Cambridge, MA 02138, USA\\
$^{11}$TAPIR, Mailcode 350-17, California Institute of Technology, Pasadena, CA 91125, USA
}

\date{Accepted XXX. Received YYY; in original form ZZZ}

\pubyear{2024}

\begin{document}
\label{firstpage}
\pagerange{\pageref{firstpage}--\pageref{lastpage}}
\maketitle

\begin{abstract}
We investigate the formation of dense stellar clumps in a suite of high-resolution cosmological zoom-in simulations of a massive, star forming galaxy at $z \sim 2$ under the presence of strong quasar winds.  Our simulations include multi-phase ISM physics from the Feedback In Realistic Environments (FIRE) project and a novel implementation of hyper-refined accretion disk winds. We show that powerful quasar winds can have a global negative impact on galaxy growth while in the strongest cases triggering the formation of an off-center clump with stellar mass ${\rm M}_{\star}\sim 10^{7}\,\Msun$, effective radius ${\rm R}_{\rm 1/2\,\rm Clump}\sim 20\,{\rm pc}$, and surface density $\Sigma_{\star} \sim 10^{4}\,\Msunpc$. The clump progenitor gas cloud is originally not star-forming, but strong ram pressure gradients driven by the quasar winds (orders of magnitude stronger than experienced in the absence of winds) lead to rapid compression and subsequent conversion of gas into stars at densities much higher than the average density of star-forming gas.  The AGN-triggered star-forming clump reaches ${\rm SFR} \sim 50\,\Msunyr$ and $\Sigma_{\rm SFR} \sim  10^{4}\,\Msunyrkpc$, converting most of the progenitor gas cloud into stars in  $\sim$2\,Myr, significantly faster than its initial free-fall time and with stellar feedback unable to stop star formation. In contrast, the same gas cloud in the absence of quasar winds forms stars over a much longer period of time ($\sim$35\,Myr), at lower densities, and losing spatial coherency.  The presence of young, ultra-dense, gravitationally bound stellar clumps in recently quenched galaxies could thus indicate local positive feedback acting alongside the strong negative impact of powerful quasar winds, providing a plausible formation scenario for globular clusters.
\end{abstract}




\section{Introduction} \label{sec:intro}
A broad range of galaxy formation models suggest that feedback from accreting supermassive black holes (BHs) in the core of active galaxies, also known as Active Galactic Nuclei (AGN), plays a critical role in the evolution of galaxies and is likely responsible for a variety of observed phenomena \citep{Hopkins2010,AlexanderHickox2012,Somerville&Dave2015,harrison2018,DiMatteo2023}. AGN feedback manifests in different forms operating on varying scales, with examples including fast accretion-driven winds \citep{Faucher-Giguere2012,Faucher-Giguere2012_model,Zubovas2012a,Tombesi2013,Nardini2015}, galaxy scale outflows \citep{feruglio2010,sturm2011,Greene2012,cicone2014,Zakamska2014,Circosta2018,Wylezalek2020,RamosAlmeida2022}, and large-scale jets \citep{Fabian2012}.
Observed strong winds powered by luminous AGN \citep[][]{Alatalo2015,Wylezalek2016,Fiore2017,Harrison2017,Wylezalek2020} can potentially provide the {\it negative} effects required in galaxy evolution models to reduce the star formation rate (SFR) in massive galaxies, but despite much recent progress, the detailed propagation and impact of AGN winds from parsec (pc) to circumgalactic medium (CGM) scales is still not fully understood \citep{Somerville&Dave2015,Hopkins2016,harrison2018,Choi2018,Costa2020,Torrey2020,Byrne2023formation,DiMatteo2023,Wellons2023}. 

In contrast, some observations suggest that AGN feedback can have {\it positive} effects, triggering  rather than suppressing star formation in galaxies. Plausible signatures of positive AGN feedback include the identification of ongoing star formation in outflowing material \citep{Santoro2016,maiolino2017,cresci2018,gallagher2019,RodriguezdelPino2019}, the spatial anti-correlation between wind-dominated central cavities and high star-forming regions \citep{cresci2015a,cresci2015b,Carniani2016,shin2019,Perna2020,Bessiere2022,Schutte2022}, jet-induced star formation within the host-galaxy \citep{Bicknell2000,Zirm2005,Drouart2016}, and large-scale bubbles driven by jets possibly triggering star formation in other galaxies \citep{Gilli2019}. 
In some cases, spatially resolved observations seem to indicate that 
positive and negative AGN feedback can coexist and operate simultaneously within a single host galaxy \citep{cresci2015b,AlYazeedi2021,Bessiere2022}.

Some idealized simulations and analytic models have proposed that positive triggering of star formation could be the dominant outcome of AGN feedback, with several works arguing that positive AGN feedback can explain the similarity in the cosmic history of star formation and AGN activity, trigger observed extreme starbursts in high-redshift galaxies, or even drive the BH–galaxy scaling relations \citep{Gaibler2012,ishibashi2012,Zubovas2012,silk2013,zubovas2013,nayakshin2014,bieri2015,bieri2016,zubovas2017}. These models are in stark contrast with a variety of hydrodynamic simulations of galaxy evolution in a cosmological context, where AGN feedback is implemented to negatively impact star formation in massive galaxies \citep{Choi2015,Schaye2015,Hirschmann2016,Angles-Alcazar2017a,Tremmel2017,Weinberger2017,Dave2019,Dubois2021,Habouzit2021,Habouzit2022,Byrne2023formation,Wellons2023}.
Given the difficulty in explicitly modelling the propagation and impact of AGN winds across scales in a full cosmological context \citep{Somerville&Dave2015,DiMatteo2023} and the degeneracies between sub-grid model parameters in cosmological large-volume simulations \citep{Villaescusa-Navarro2021,Jo2023_camels,Ni2023}, it has remained a challenge to fully discriminate between positive and negative AGN feedback scenarios.

In \citet{Mercedes-Feliz2023}, we investigated the plausible dual role of AGN feedback in galaxies using high-resolution cosmological zoom-in simulations from the Feedback
In Realistic Environments (FIRE\footnote{\url{http://fire.northwestern.edu}}) project \citep{Hopkins2014,Hopkins2018,Hopkins2023}, implementing local star formation and stellar feedback processes in a multi-phase interstellar medium (ISM) while also including a novel implementation of hyper-refined accretion-driven AGN winds that captures self-consistently their propagation and impact from the inner 10\,pc to CGM scales \citep{Byrne2023formation,Cochrane2023,Hopkins2023,Wellons2023,Angles-Alcazar2023}. These simulations are among the most detailed models of a powerful quasar phase in a massive star-forming galaxy at its peak of activity ($M_{\rm halo}\sim 10^{12.5}\,{\rm M}_{\odot}$ at $z=2$) and are thus ideally suited to investigate the impact of AGN winds on resolved galaxy properties. 
Comparing identical simulations with either no AGN feedback or varying AGN feedback strength, \citet{Mercedes-Feliz2023} demonstrated that strong quasar winds with kinetic power $\sim$10$^{46}$ erg/s persisting for $\sim$20\,Myr can have a strong global negative impact on the host galaxy, driving the formation of a central gas cavity and significantly reducing the SFR surface density across the galaxy disc. Nonetheless, we identified several potential indicators of local positive AGN feedback coexisting with the global negative effects, including spatial anti-correlations between wind-dominated regions and star-forming clumps similar to observations \citep{cresci2015a,Carniani2016,shin2019},  higher local star formation efficiency in compressed gas at the edge of the cavity, as seen in some local active galaxies \citep{shin2019,Perna2020,Schutte2022}, and the presence of outflowing material with ongoing star formation, qualitatively consistent with some observations \citep{maiolino2017,gallagher2019}.

In this work, we extend our previous analyses to investigate in more detail what appears to be the strongest manifestation of positive AGN feedback occurring in our simulations: the formation of very dense stellar clumps with stellar mass ${\rm M}_{\star}\sim 10^{7}\,\Msun$, stellar effective radius ${\rm R}_{\rm 1/2\,\rm Clump}\sim 20\,{\rm pc}$, and stellar surface density $\Sigma_{\star} \sim 10^{4}\,\Msunpc$.
These extreme clumps occur exclusively in our simulations with very strong AGN winds, while their presence is not observed in simulations with weaker AGN feedback. The presence of ultra-dense stellar clumps in galaxies that are otherwise experiencing global quenching of star formation could thus be a unique signature of co-existing local positive and global negative feedback by powerful quasar winds.  Here we reconstruct the full evolution of these stellar clumps in detail and demonstrate the direct role of quasar winds on their formation.

The outline of this paper is as follows: \S\ref{sec:methods} provides a brief summary of the galaxy formation framework and our methodology to implement AGN winds; \S\ref{sec:overview} presents an overview of the simulations and the identified stellar clumps; \S\ref{sec:SFunderEC} explores the impact of AGN winds on global and local star formation; \S\ref{sec:AGNdriveCF} investigates the direct role of strong AGN winds driving the formation of the stellar clumps; \S\ref{sec:discussion} discusses our results in the context of previous work; and \S\ref{sec:summary} provides a summary of our findings and the main conclusions of this work.

\begin{table}
\centering
\begin{tabular}{c| c c c c} 
    \hline \hline
    Name & $\eta_{k}$ & $\epsilon_{k}$ & $\dot{M}_{\rm w}\,[\Msunyr]$ & $\dot{E}_{\rm w}\,[{\rm erg}\, {\rm s}^{-1}]$ \\ [0.05ex] 
    \hline
    \textbf{noAGN} & - & - & - & -  \\ 
    \hline
    \textbf{m0.1e0.5} & 0.1 & 0.005  & 2.22 & $6.29 \times 10^{44}$ \\
    \hline
    \textbf{m1e5} & 1 & 0.05 & 22.2 & $6.29 \times 10^{45}$ \\
    \hline
    \textbf{m2e10} & 2 & 0.1 & 44.4 & $1.26 \times 10^{46}$ \\
    \hline
    \textbf{m4e20} & 4 & 0.2 & 88.8 & $2.52 \times 10^{46}$ \\
    \hline
    \textbf{m10e50} & 10 & 0.5  & 222 & $6.29 \times 10^{46}$ \\ [1ex] 
    \hline
    \end{tabular}
\caption{{Simulation parameters: (1) Name: simulation designation. (2) $\eta_{k} \equiv \dot{M}_{w}/\dot{M}_{\rm BH}$: mass loading factor. (3) $\epsilon_{k} \equiv \dot{E}_{\rm w}/L_{\rm bol}$: kinetic feedback efficiency. (4) $\dot{M}_{w}$: mass outflow rate in winds. (5) $\dot{E}_{w}$: kinetic energy injection rate.}}
\label{table:models}
\end{table}

\section{Methods} \label{sec:methods}

The simulations and methodology that we use are presented and fully described in \citet{Angles-Alcazar2023}, which we briefly summarize below. 
The same simulations have been previously analyzed in \citet{Mercedes-Feliz2023} and \citet{Cochrane2023}.

\subsection{FIRE-2 galaxy formation model} \label{subsec:FIRE2model}
Our simulations are part of the Feedback In Realistic Environments (FIRE) project\footnote{\url{http://fire.northwestern.edu}} and we use specifically the ``FIRE-2'' galaxy formation physics implementation \citep{Hopkins2018}. The simulations use the $N$-body and hydrodynamics code GIZMO\footnote{\url{http://www.tapir.caltech.edu/~phopkins/Site/GIZMO.html}} in its ``meshless finite mass'' (MFM) hydrodynamics mode \citep{Hopkins2015gizmo}, a Lagrangian Godunov formulation which sets both hydrodynamic and gravitational (force-softening) spatial resolution in a fully-adaptive Lagrangian manner, with fixed mass resolution. As outlined in \citet{Hopkins2018}, we include cooling and heating from $T=10-10^{10}\,{\rm K}$; star formation in locally self-gravitating, dense ($n_{\rm H}\geq n_{\rm H, th} \equiv 1000\,{\rm cm}^{-3}$), molecular, and Jeans-unstable gas; and stellar feedback from OB \&\ AGB mass-loss, Type Ia \&\ II Supernovae (SNe), and multi-wavelength photo-heating and radiation pressure; with each star particle representing a single stellar population with known mass, age, and metallicity with all stellar feedback quantities and their time dependence directly taken from the \textsc{starburst99} population synthesis model \citep{Leitherer1999}.

\subsection{Initial conditions} \label{subsec:initialconditions}

Our simulations use snapshots from pre-existing FIRE-2 simulations as initial conditions to perform new simulations that include AGN-driven winds. We focus primarily on the massive FIRE-2 halo \textbf{A4} from \citet{Angles-Alcazar2017c}. In those simulations, the halo reaches mass $M_{\rm halo}\sim 10^{12.5}\,{\rm M}_{\odot}$ at $z=2$ and was evolved down to $z=1$ including on-the-fly BH growth driven by gravitational torques \citep{Hopkins&Quataert2011,Angles-Alcazar2013,Angles-Alcazar2015,Angles-Alcazar2017a} but no AGN feedback. The new simulations with AGN winds adopt the same baryonic (gas and stellar) mass resolution $m_{\rm b}=3.3\times 10^{4}\,{\rm M}_{\odot}$ and dark matter mass resolution $m_{\rm DM}=1.7\times 10^{5}\,\Msun$ as well as gravitational force softenings $\epsilon_{\rm gas}^{\rm min}=0.7\,{\rm pc}$, $\epsilon_{\star}=7\,{\rm pc}$ and $\epsilon_{\rm DM}=57\,{\rm pc}$ for the gas (minimum adaptive force softening), stellar, and dark matter components. We assume a $\Lambda$CDM cosmology with parameters $H_{0}=69.7\,{\rm km}\,{\rm s}^{-1}\,{\rm Mpc}^{-1}$, $\Omega_{\rm M}=1-\Omega_{\Lambda}=0.2821$, $\Omega_{\rm b}=0.0461$, $\sigma_{8}=0.817$, and $n_{\rm s}=0.9646$ \citep{Hinshaw2013}.

We select the $z=2.28$ simulation snapshot as the time to inject AGN winds in the new simulations, which will be referenced as $t_{0}\equiv\Delta t=0\,{\rm Myr}$ throughout the rest of the paper. At this time, the galaxy is undergoing a strong starburst phase which will lead to the formation of an overcompact and overdense stellar component due to stellar feedback no longer being able to regulate star formation  \citep{Wellons2020,Parsotan2021,Cochrane2023,Angles-Alcazar2023}. A separate set of Lagrangian hyper-refinement simulations have shown explicitly that strong gravitational torques from the stellar component are driving at this time an inflow rate down to sub-pc scales sufficient to power a luminous quasar \citep{Angles-Alcazar2021}, motivating further our choice of initial conditions to re-simulate including strong quasar winds.  
For further details, see \citet{Mercedes-Feliz2023}.

\begin{figure*}
\includegraphics[width = \textwidth]{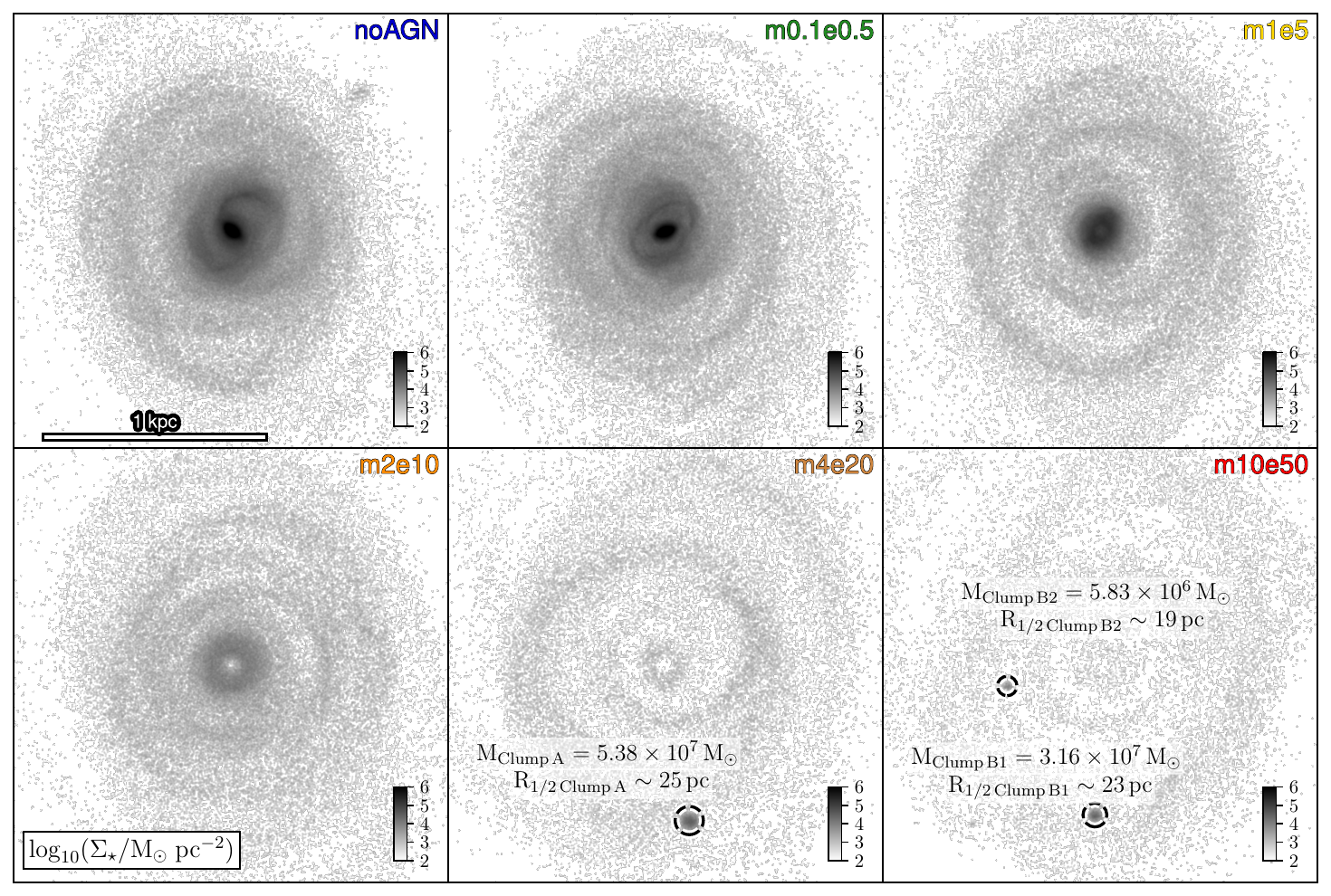}
\vspace*{-5mm}
\caption{Projected stellar mass surface density for the central 1\,kpc region of a massive, star-forming galaxy ($M_{\rm star}\sim 10^{11}\,\Msun$, ${\rm SFR}\sim 300\,\Msunyr$) at $z\sim2.28$ for various AGN feedback efficiencies after $\sim 35\,{\rm Myr}$ since the start ($\Delta t = 0$) of the quasar wind phase. From top left (no AGN feedback) to bottom right (strongest AGN winds), we show face-on views of the population of stars that formed in the last $\sim 35\,{\rm Myr}$ in simulations with increasing AGN feedback strength. We identify the presence of dense stellar clumps that only form in the two strongest AGN feedback cases despite their overall suppression of star formation.}
\label{fig:clusters_all} 
\end{figure*}

\begin{figure*}
\includegraphics[width = \textwidth]{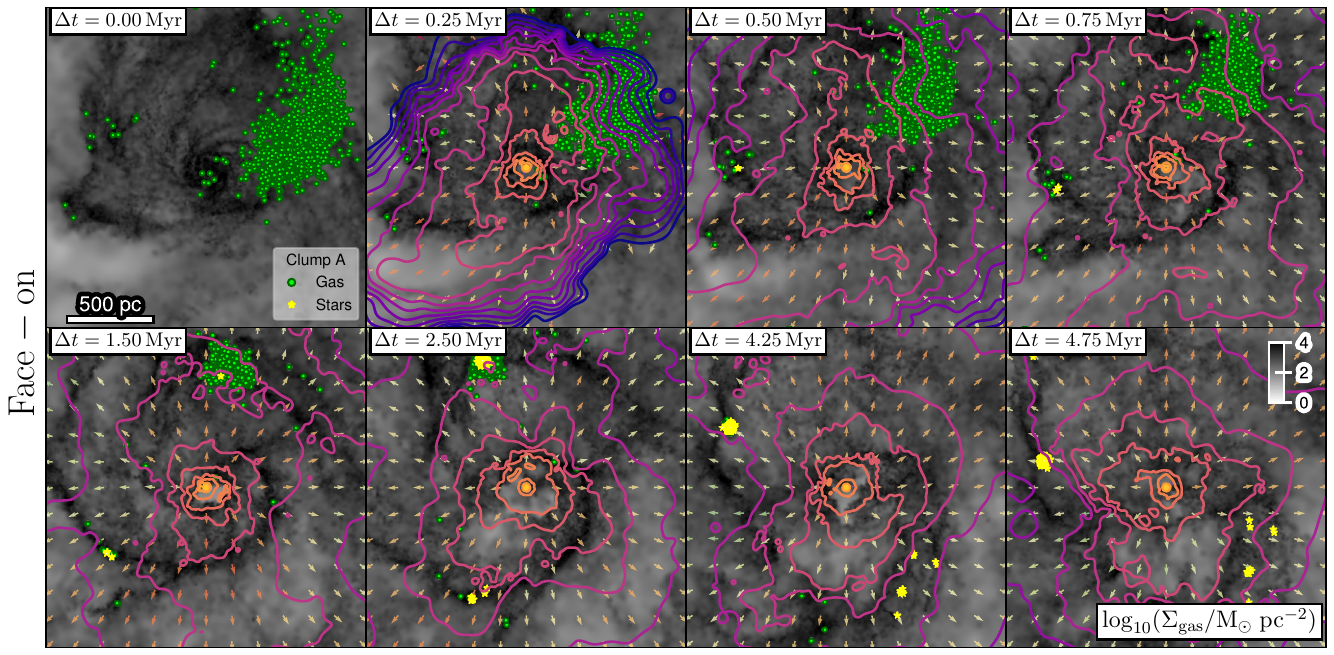}
\includegraphics[width = \textwidth]{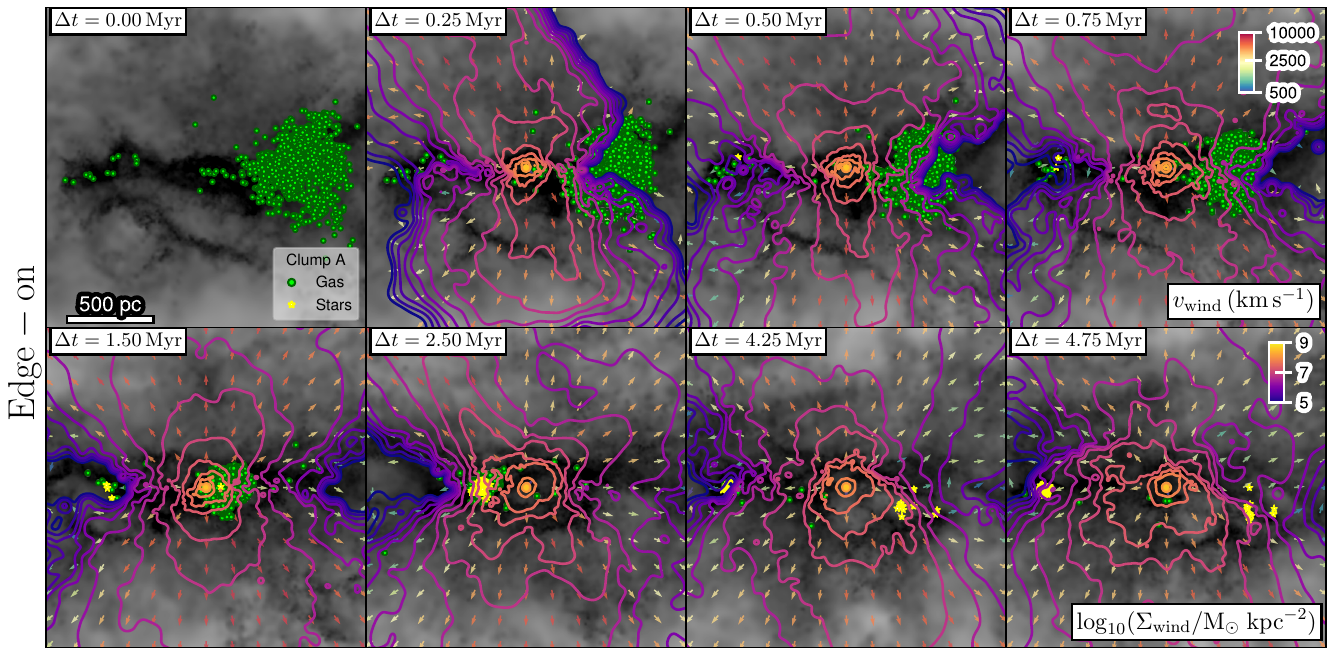}
\vspace*{-5mm}
\caption{Gas mass surface density maps (grey scale) at eight different times during the formation of Clump A in simulation \mf, for the face-on (top two rows) and edge-on (bottom two rows) views of the central $(2\,{\rm kpc})^2$. The particles that end up forming the clump are shown in their gas phase (green points) until they turn into stars (yellow points). The contours outline the AGN wind mass surface density, while the vector field denotes the AGN wind velocity field. The gas cloud, originally spread over $\sim 1$\,kpc, is compressed by the AGN winds radially through the galaxy disk and vertically by the expanding bi-conical outflow.}
\label{fig:clusterA_formingFOEO} 
\end{figure*}

\subsection{Hyper-refined black hole-driven winds} \label{subsec:BHdrivenwinds}

We inject AGN winds at hyper-Lagrangian resolution using the method described in \citet{Angles-Alcazar2023}. This method builds on earlier particle spawning techniques in idealized simulations of galaxies and massive haloes \citep{Richings2018,Torrey2020,Su2021} and has now been implemented in FIRE-3 simulations with BH physics \citep{Hopkins2023,Wellons2023}. The BH is modelled as a collisionless particle with an initial mass $M_{\rm BH}=10^{9}\,\Msun$, located at the center of the main simulated galaxy. 
Since the BH mass is much larger than the baryonic and dark matter particle masses, the BH dynamics is fully resolved and we do not need to artificially force the BH to stay at the center of the galaxy. For simplicity, the BH is assumed to be accreting at a constant rate throughout the entire simulation, set at the Eddington rate ($\dot{M}_{\rm BH}=22.2\,\Msunyr$), which represents a luminous quasar phase ($L_{\rm bol}\sim 10^{47}\,{\rm erg}\,{\rm s}^{-1}$) continuously powering winds for $\sim$40\,Myr.
Mass conservation is ensured with stochastic swallowing of gas particles within the BH interaction kernel \citep[defined to contain $\sim 256$ particles, e.g.,][]{Angles-Alcazar2017a}.

The following main properties specify our AGN wind model: the mass outflow rate $\dot{M}_{\rm w}$, the initial wind velocity $v_{\rm w}$, and the initial wind geometry. We consider that a fraction $\epsilon_{\rm k}$ of the AGN bolometric luminosity emerges as a fast isotropic wind that expands radially outward from the BH, with an initial velocity $v_{\rm w} = 30,000\,{\rm km}\,{\rm s}^{-1}$ and temperature $T_{\rm w}\sim 10^{4}\,{\rm K}$. We assume that the wind immediately interacts with the ambient medium and attains a post-shock velocity and temperature given by $v_{\rm sh} =v_{\rm w}/4 = 7,500\,\kms$ and $T_{\rm sh} \approx 1.2\times 10^{10}$\,K \citep{Faucher-Giguere2012}. We model the AGN wind by spawning new gas particles within a sphere $R_{\rm w}=0.1\,{\rm pc}$ around the BH, with an initial velocity and temperature given by the post-shock properties $v_{\rm sh}$ and $T_{\rm sh}$, implementing a target wind particle mass of $1000\,\Msunh$ ($>$20 times higher mass resolution than the original simulation). We implement discrete ejection events containing $N_{\rm w}=10-100$ wind particles distributed isotropically and moving radially outward from the BH. 
The total gas mass accreted into the BH ($\Delta M_{\rm BH}$) and the total mass of winds injected into the simulation ($\Delta M_{\rm w}$) are calculated at each timestep, where the combined accreted and spawned gas mass is removed from pre-existing gas to satisfy mass conservation in the simulation. 
Wind injection events occur frequent enough to appear quasi-continuously but always with enough particles to represent an isotropic wind (our results are not sensitive to $N_{\rm w}$). Other fluid quantities are immediately recomputed for the wind particles after spawning, modelling self-consistently the hydrodynamic interaction of winds with the ISM gas of the host galaxy. Other than their mass, wind particles are thus treated identically to preexisting gas in the simulation and can even participate in star formation (though this rarely happens due to wind particles generally not satisfying the criteria for star formation).
Once the wind particles slow down and reach a velocity lower than 10\%~of the initial wind velocity ($<750\,\kms$), they are allowed to merge with the nearest gas element to reduce the computational cost of the simulation. Wind particles at this point have transferred most of their energy and momentum to the surrounding gas and further following their evolution alongside regular gas particles becomes less relevant. Particle spawning allows us to fully capture the propagation and impact of fast winds with Lagrangian hyper-refinement \citep[see also][]{Richings2018,Torrey2020,Costa2020}, injecting feedback locally around the BH and capturing the wind-ISM interaction robustly regardless of gas geometry and at significantly higher resolution than nearest neighbor-based feedback coupling models. 

Table \ref{table:models} summarizes the main properties of the simulations analyzed here. All simulations start from the same initial conditions described in $\mathsection$\ref{subsec:initialconditions}, containing a central BH with mass $M_{\rm BH}=10^{9}\,{\rm M}_{\rm \odot}$ accreting at the Eddington rate, and implementing the same post-shock wind velocity and temperature while varying the mass outflow rate $\dot{M}_{\rm w}$. Along with the standard FIRE-2 simulation
that excludes AGN feedback (\nf), we investigate the impact of AGN-driven winds with kinetic feedback efficiencies in the range $\epsilon_{\rm k} =0.5$–50\%, which brackets a range of observational constraints \citep[e.g.][]{cicone2014,Fiore2017,harrison2018} and assumed feedback efficiencies in previous simulations \citep[e.g.][]{DiMatteo2005,Weinberger2017,Dave2019}. The simulation name in each feedback case encodes the value of the mass loading factor ($\eta_{\rm k}$) and the kinetic feedback efficiency ($\epsilon_{\rm k}\times 100$).
Our choice in BH mass and accretion rate is representative of those found in luminous quasars at $z\sim2$ given the host galaxy stellar mass ($\sim 10^{11}\,\Msun$; e.g., \citealt{Trakhtenbrot2014}; \citealt{Zakamska2019}). However, the assumed AGN wind kinetic efficiency exhibits a degeneracy with the chosen BH mass and Eddington ratio. For example, by selecting a BH mass or Eddington ratio a factor of 10 lower and simultaneously increasing the kinetic efficiency by a factor of 10, we would achieve equivalent mass, momentum, and energy injection rates for the resulting AGN winds (which are the actual relevant physical parameters in the simulations presented here).
The times mentioned in this work are relative to the start of the quasar phase at $t_{0}$, with $\Delta t$ referring to the time that has passed since then as $\Delta t\equiv t-t_{0}$. The two simulations that we reference the most throughout this work are:
\begin{enumerate}[wide, labelwidth=!,itemindent=!]
    \item \nf: The control simulation using standard FIRE-2 physics, where we model the evolution of a massive galaxy ($M_{\rm star}\sim 10^{11}\,\Msun$) starting at $z\sim 2.28$ ($t_{0} \equiv \Delta t =0$\,Myr) and no AGN winds are introduced. The BH is still accreting at the Eddington rate, $\dot{\rm M}_{\rm BH} \sim 22.2\,\Msunyr$.\\
    \item \mf: AGN winds are turned on at $\Delta t =0$\,Myr with the same initial conditions as the \nf~case. We consider a luminous quasar phase with bolometric luminosity $L_{\rm bol} = 1.26\times 10^{47}$\,erg\,s$^{-1}$, driving a wind with kinetic efficiency $\epsilon_{\rm k}=0.2$ and mass loading factor $\eta_{\rm k}\equiv \dot{M}_{w}/\dot{M}_{\rm BH}= 4$, corresponding to a mass outflow rate in winds $\dot{M}_{w}=88.8\,\Msunyr$.
\end{enumerate}
The simulations listed in Table \ref{table:models} are evolved for different lengths of time ($\sim$35--70\,Myr) but we focus on the first $\Delta t \sim 35\,{\rm Myr}$ of evolution throughout this work, which we refer to as the end of the simulated quasar phase. We save snapshots every $0.01\,{\rm Myr}$ for $0\leq\Delta t<5\,{\rm Myr}$ and every $0.1\,{\rm Myr}$ for $\Delta t> 5\,{\rm Myr}$ for all of the AGN-wind simulation, while the time between snapshots in the \nf~case is $0.2\,{\rm Myr}$ for the duration of the simulation.

\begin{figure*}
\includegraphics[width = \textwidth ]{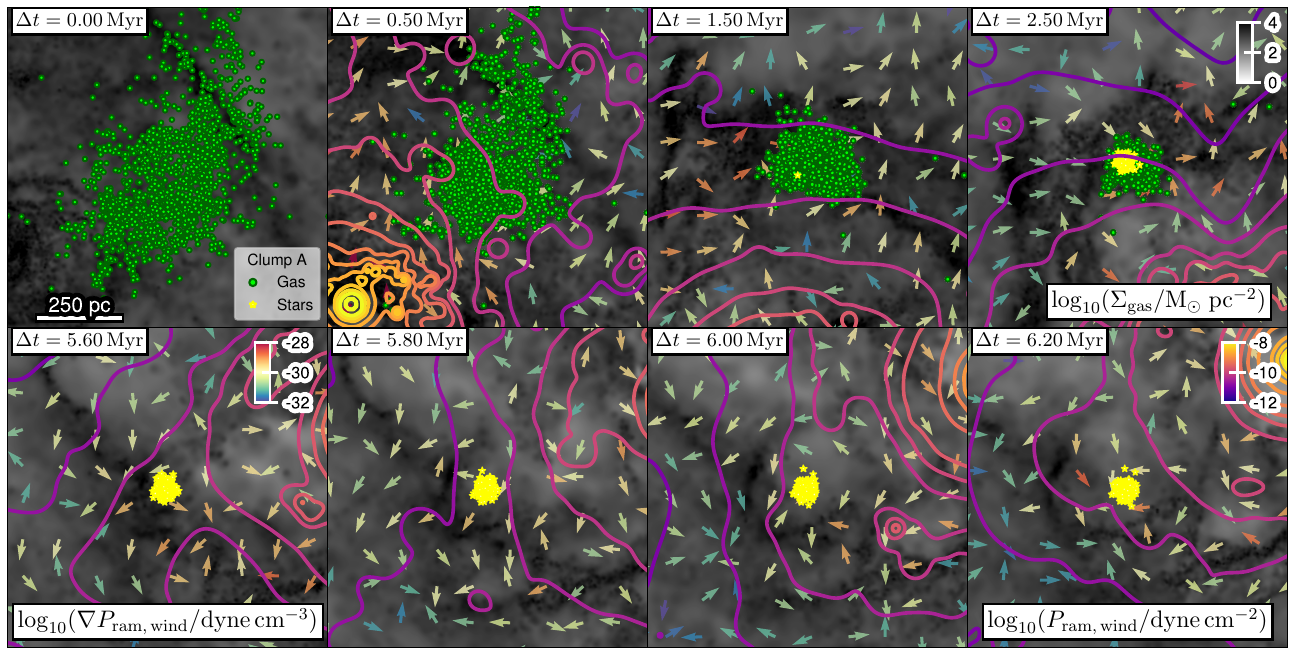}
\vspace*{-3mm}
\caption{Similar to Figure \ref{fig:clusterA_formingFOEO} except focusing on $(1\,{\rm kpc})^2$ face-on projected gas surface density maps (grey scale) centered on Clump A. The contours outline the mass-weighted ram pressure of the AGN wind while the vector field denotes the direction of the force exerted by the ram pressure gradient (color-coded by the magnitude of the ram pressure gradient). The AGN wind ram pressure compression drives the formation of an ultra-dense gas clump quickly converting $\sim 5\times 10^{7}\,\Msun$ into stars, with stellar feedback unable to regulate star formation and only driving a quasi-spherical expanding gas shell at $\Delta t \sim 6\,{\rm Myr}$ after the stellar clump has fully formed.}
\label{fig:clusterA_forming} 
\end{figure*}

\section{Overview of Simulations} \label{sec:overview}
Figure \ref{fig:clusters_all} shows the face-on projected stellar mass surface density for six different simulations of the same massive star-forming galaxy ($M_{\rm star}\sim10^{11}\,{\rm M}_{\odot}$, ${\rm SFR}\sim300\, {\rm M}_{\odot}\, {\rm yr}^{-1}$ at $z\sim 2.28$), one with no AGN feedback (\nf) and the rest including AGN feedback, with their AGN wind parameters varied as shown in Table \ref{table:models}. Each panel shows the population of stars that formed in the $\Delta t=35\,{\rm Myr}$ since the start of the quasar wind phase ($\Delta t=0$). The \nf~simulation shows that, in the absence of AGN winds, the galaxy forms an ultra-dense nuclear stellar disk, with ${\rm M}_{\star}\sim \times 10^{10}\,{\rm M}_{\odot}$ in the central $100$\,pc and nuclear spiral and bar-like features. The second panel corresponds to the simulation with the weakest AGN winds, \simone. With a mass outflow rate of $\dot{{\rm M}}_{w}\sim 2.22\,\Msunyr$ and a kinetic efficiency $\epsilon_{\rm k}=0.005$, AGN feedback only decreases slightly the stellar mass formed relative to the \nf~simulation. As we continue to increase the strength of AGN winds with $\dot{{\rm M}}_{w}=22.2\,\Msunyr$ and $\epsilon_{\rm k}=0.05$ in simulation \simtwo, winds are strong enough to create a nuclear gas cavity ejecting a considerable amount of gas from the center, reducing star formation and the total stellar mass within $100$\,pc to ${\rm M}_{\star}\sim 2.5\times 10^{9}\,{\rm M}_{\odot}$. With the reduced global SFR, we begin to see ring-like structures with stars tracing the spiral arms where they formed. For the first panel in the bottom row, \simthree, we now have an outflow rate of $\dot{{\rm M}}_{w}=44.4\,\Msunyr$, where the gas cavity evacuated by the winds reached $\sim 300$\,pc \citep{Mercedes-Feliz2023}, leaving a cavity imprinted also in the stellar component, and the global suppression of star formation limits the total stellar mass growth to ${\rm M}_{\star}=2.32\times 10^{9}\,{\rm M}_{\odot}$.

The second strongest feedback case, \simfour, with a wind outflow rate of $\dot{{\rm M}}_{w}=88.8\,\Msunyr$, has a dramatic negative impact on the global SFR, limiting the stellar mass growth to ${\rm M}_{\star}\sim 1\times 10^{9}\,{\rm M}_{\odot}$. In this case, besides the central cavity in the stellar distribution and the ring-like structures, the most prominent feature is a very dense stellar clump located $\sim 700$\,pc from the center of the galaxy. This dense stellar region, which we denote as Clump A, has a half mass radius of ${\rm R}_{1/2 \rm \,Clump\,A}\sim 25\,{\rm pc}$ with an enclosed stellar mass of ${\rm M}_{\rm Clump\,A}=5.38\times10^{7}\,{\rm M}_{\odot}$. The strongest feedback case, \simfive, with an outflow rate of $\dot{{\rm M}}_{w}=222\,\Msunyr$, has an even more dramatic effect on the galaxy, with the total stellar mass growth limited to ${\rm M}_{\star}\sim 6 \times 10^{8}\,{\rm M}_{\odot}$. Interestingly, despite the strong suppression of star formation, we find two very dense stellar clumps with properties similar to Clump A in \mf. We denote these two dense stellar regions as Clump B1 and Clump B2, with similar half mass radii ${\rm R}_{1/2 \rm \,Clump\, B1}\sim 23\,{\rm pc}$ and ${\rm R}_{1/2 \rm \,Clump\, B2}\sim 19\,{\rm pc}$ and stellar masses ${\rm M}_{\rm Clump\,B1}=3.16\times10^{7}\,{\rm M}_{\odot}$ and ${\rm M}_{\rm Clump\,B2}=5.83\times10^{6}\,{\rm M}_{\odot}$, respectively. 
In this paper, we analyze in detail the formation of these ultra dense stellar clumps driven by strong AGN winds.

Clumps are first visually identified in stellar surface density maps corresponding to newly formed stars in the last $\Delta t = 35\,{\rm Myr}$ (Figure \ref{fig:clusters_all}). 
We then compute the center of mass of all newly formed stars within a radial aperture of 100\,pc and identify as clump members the star particles within a 3D radius enclosing 95\%~of the mass within 100\,pc. Throughout this paper, clump sizes and masses are computed based on these selected clump particle members but our main results are not sensitive to this choice.
Figure \ref{fig:clusterA_formingFOEO} illustrates the formation of Clump A in simulation \mf, where we show the projected gas surface density distribution for the central $(2\,{\rm kpc})^3$ region at various snapshots in time. Green points indicate the location of gas particles that end up forming stars in Clump A (switching to yellow as they turn into stars) and the contours represent the wind mass surface density (\sigmawind) with the arrows indicating the wind velocity. The top two rows show a face-on view of how the clump particles as well as the surrounding ISM interact with the AGN-driven winds, while the bottom two rows show an edge-on view to better indicate the depth at which the winds penetrate through the galaxy. At the beginning of the simulation, the galaxy resembles a turbulent, clumpy, kpc-scale disk with dense gas regions along fractured spiral arms and the even denser gas within the nuclear region. Progenitor Clump A particles are initially spread over a kpc-scale region, encompassing gas structures in two spiral arms and extending beyond the plane of the disk.  At $\Delta t=0.25$\,Myr, the AGN winds have effectively reached $>1$\,kpc whilst pushing the clump particles radially outwards in the plane of the disk and squeezing them into a wedge in the vertical direction. As the simulation proceeds, the AGN winds continue to propagate outward, blowing out the gas closest to the SMBH while compressing the progenitor gas particles (a significant fraction still infalling) into a dense gas clump that quickly turns into stars.

To further study the formation of Clump A, in simulation \mf, Figure \ref{fig:clusterA_forming} provides an in-depth examination by showing face-on gas surface density projections, zooming into the 1\,kpc region around the center of mass of Clump A. The contours are now outlining the mass-weighted ram pressure for the AGN winds, while the vector field shows the ram pressure
gradient, in order to find any correlations with the clump formation. We achieve this by computing the ram pressure (see \S\ref{subsec:timeevolution} for more details) and its gradient considering all gas particles, including the AGN winds as they interact hydrodynamically with the ISM gas. However, we show the mass-weighted ram pressure and gradient for the AGN wind particles alone to highlight their effect on the formation of the clump. The top row shows how the AGN winds have impacted the clump particles as they continue to push the inner parts of the cloud while the outer part continues to infall to the center of the gravitational potential well, further compressing the cloud. The lower row focuses on the clump once it has efficiently exhausted its gas content, quickly forming into stars before stellar feedback can regulate star formation. We also see how SNe remove any remaining gas as a quasi-spherical shell expanding in the last three panels. 

Every particle in the simulation has a unique set of identifiers (IDs), which are the same across snapshots, allowing us to link particles across time. This is useful for tracking particles between snapshots or even between the different simulation suites. Once a gas particle satisfies the star-forming conditions briefly mentioned in \S\ref{subsec:FIRE2model}, it turns into a star particle. That star particle will retain the IDs of its ``parent'' gas particle. This allows us to find and trace star particles back in time to before they formed. We select all star particles that have formed within $\sim$35\,Myr since the beginning of the quasar phase for each simulation (as seen in Figure~\ref{fig:clusters_all}) and track them back in time to the start of the quasar wind phase ($\Delta t=0$), identifying their last instance as gas particle and recording the ``final'' gas density ($n_{\rm H}$) right before turning into a star particle.
Figure \ref{fig:35MyrallSIMS} shows the normalized probability distribution of final gas densities for stars that formed since the start of the quasar phase for each simulation. The black dotted line highlights the peak in the \nf~(blue) distribution. By using the \nf~simulation as the baseline, we see two interesting trends in the densities that gas particles reach right before turning into stars as we vary feedback parameters. In the simulations with relatively weak feedback, \simone~(green) and \simtwo~(yellow), a larger number of stars formed at densities higher than ${\rm log}_{10}(n_{\rm H}/\,{\rm cm}^{-3})=4.5$, with the peak skewing to higher densities as we make the AGN winds stronger. For simulations with even stronger winds, however, we see the opposite trend, with more stars forming at lower densities, the peak of the distribution shifting to ${\rm log}_{10}(n_{\rm H}/\,{\rm cm}^{-3})=4$, and a tail end extending to densities as high as $n_{\rm H}\sim 10^{6}\,{\rm cm}^{-3}$. We will explicitly compare the same set of particles tracked across simulations below, but we can already see that varying the AGN wind strength can change the densities at which stars are forming for the overall stellar population. 

\section{Star Formation under Extreme Conditions}\label{sec:SFunderEC}

\subsection{Gas density prior to star formation}\label{subsec:beforeSF}
In this section, we make comparisons between star-forming particles within the \mf~simulation and those of the Clump A particles. We compare various properties in order to identify any systematic differences in the physical conditions of gas that forms stars in dense clumps compared to the galaxy disk.

\begin{figure}
\includegraphics[width = \columnwidth ]{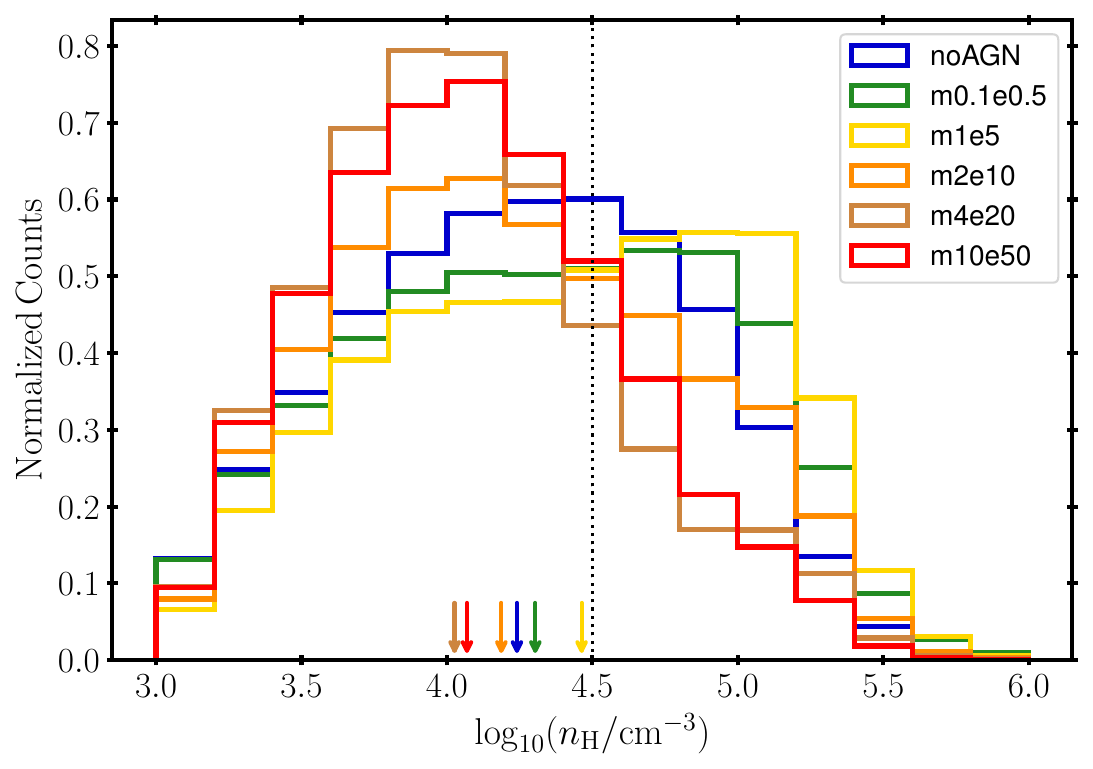}
\vspace*{-4mm}
\caption{Normalized probability distribution of the density ($n_{\rm H}$) of gas resolution elements just before being converted into stars in simulations with different AGN feedback strength, for all the stars that formed within $\sim$35\,Myr of the start of the quasar phase. The black dotted line indicates the peak of the \nf~simulation distribution, with the arrows showing the median values for each simulation. We find that stars that formed in the \nf~simulation roughly follow a symmetrical distribution around $n_{\rm H}\sim 10^{4.5}\,{\rm cm}^{-3}$, while simulations with AGN feedback can bias the distribution to either higher densities for the weaker AGN winds (\simone, green; \simtwo, yellow) or lower densities in the stronger AGN wind cases (\simthree, orange; \mf, brown; \simfive, red).}
\label{fig:35MyrallSIMS}
\end{figure}

\begin{figure}
\includegraphics[width = \columnwidth ]{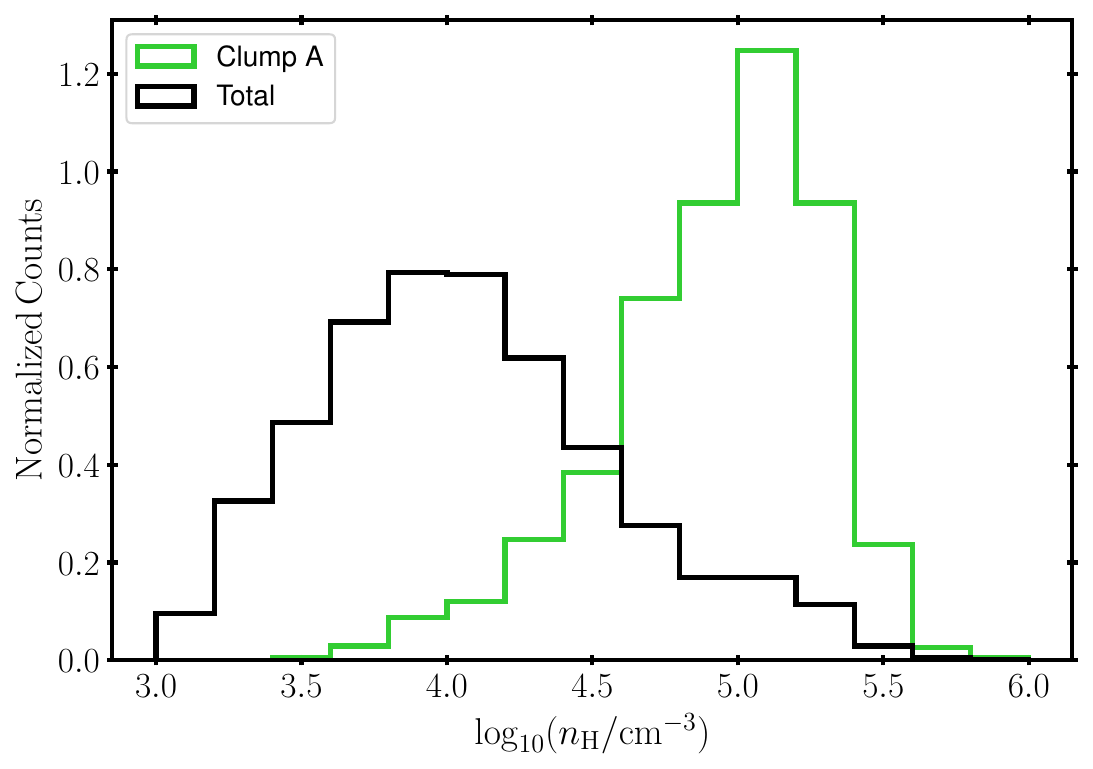}
\vspace*{-4mm}
\caption{Normalized probability distribution of the gas density ($n_{\rm H}$) at which star particles form in the \mf~simulation within $\sim 35\,{\rm Myr}$ since the start of the quasar phase (black) and shown separately for stars formed in Clump A (green). Stars in Clump A formed at significantly higher densities than the overall population of stars that form in the presence of winds.}
\label{fig:clusters_e1f02_nH} 
\end{figure}

\begin{figure}
\includegraphics[width = \columnwidth ]{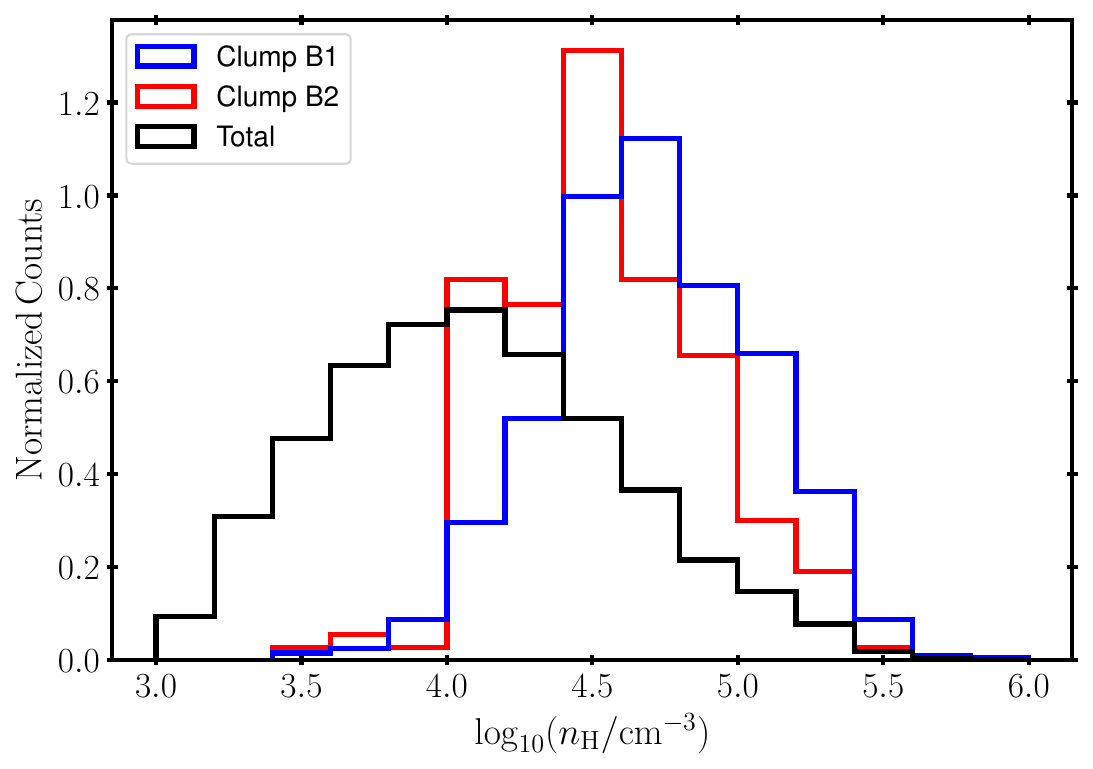}
\vspace*{-4mm}
\caption{Similar to Figure \ref{fig:clusters_e1f02_nH} but for the strongest AGN feedback case, \simfive, with the gas density distributions prior to star formation indicated for Clumps B1 (blue) and B2 (red). The clump star particles form at higher densities than the overall population of stars that form in the presence of winds, similar to Clump A in simulation \mf.}
\label{fig:clusters_e1f009_nH} 
\end{figure}

We use similar analysis as in Figure \ref{fig:35MyrallSIMS} to construct Figure \ref{fig:clusters_e1f02_nH}, where we show the normalized probability distribution of final gas densities ($n_{\rm H}$) for stars that formed within $\sim$35\,Myr in simulation \mf~(see also the middle bottom panel in Figure \ref{fig:clusters_all}), including Clump A. The black line is the distribution for all of the stars that formed within 35\,Myr of the start of the quasar phase ($\sim$35,000 particles).
As expected, all stars form at densities higher than the star formation density threshold in FIRE-2 ($n_{\rm H,th} \equiv 1000\,{\rm cm}^{-3}$), where gas is also required to be molecular, self-gravitating, and Jeans unstable to form stars \citep{Hopkins2018}.
The distribution peaks at $10^4\,{\rm cm}^{-3}$ with a long tail end to even higher densities. Alongside the overall population of stars formed within the simulation, we also show the distribution for the star particles in Clump A as the green line ($\sim$1,700 tracked particles), which is much narrower and reaches significantly higher densities than the overall stellar population. The Clump A distribution peaks at a gas density of $\sim 10^{5}\,{\rm cm}^{-3}$ with a tail end at lower densities, reaching $\sim 10^{3.5}\,{\rm cm}^{-3}$, a factor of three times above the density threshold. This indicates that Clump A particles form under quantitatively different conditions compared to the overall population of stars.

We perform the same analysis for the two clumps that we identify in simulation \simfive, which represents the strongest feedback case. Figure \ref{fig:clusters_e1f009_nH} shows the distributions of gas density prior to star formation for Clumps B1 (blue; $\sim$1,000 particles) and B2 (red, $\sim$200 particles) compared to the overall stellar population formed within 35\,Myr in the presence of AGN winds (black; $\sim$18,000 star particles). The distribution for the overall population is similar to that of the simulation \mf, with the peak at $\sim 10^{4}\,{\rm cm}^{-3}$ but with fewer stars forming overall than in the \mf~case, reflective of the impact of stronger winds and faster quenching of star formation (the \simfive~simulation forms approximately half as many stars as the \mf~simulation, in the same $35\,{\rm Myr}$ period). Interestingly, Clumps B1 and B2 combined represent about the same fraction of stars forming at $n_{\rm H}>10^{4}\,{\rm cm}^{-3}$ as Clump A relative to the overall stellar population (~$\sim$50\,\%). However, the $n_{\rm H}$ distributions for Clumps B1 and B2 differ from that of Clump A in some respects. Compared to Clump A, Clumps B1 and B2 exhibit a narrower range of densities and no prominent tail ends at either extreme. The distribution for Clump B1 peaks at $n_{\rm H} \sim 10^{4.7}\,{\rm cm}^{-3}$, while the second clump peaks at a slightly lower density $n_{\rm H} \sim 10^{4.5}\,{\rm cm}^{-3}$ (compared to $n_{\rm H} \sim 10^{5}\,{\rm cm}^{-3}$ for Clump A). In conclusion, the three distinct clumps identified in the two simulations with the strongest AGN winds form their stars at higher densities than the overall population of stars formed in the presence of AGN winds. In the following, we focus our analysis on the formation of Clump A but similar results are obtained for clumps B1 and B2.  

\subsection{Time evolution of density and pressure gradients}\label{subsec:timeevolution}
\begin{figure}
\includegraphics[width = \columnwidth, valign=t]{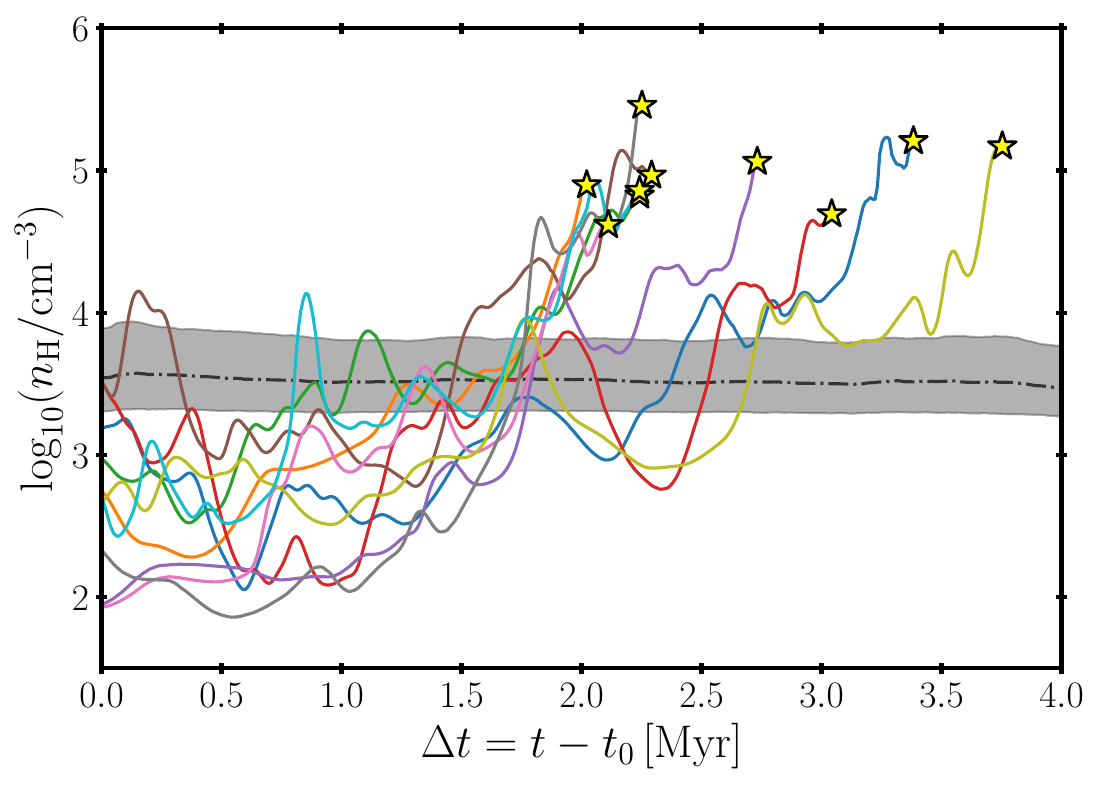}
\vspace*{-4mm}
\caption{Density ($n_{\rm H}$) of the gas particle progenitors of ten star particles randomly selected from Clump A as a function of time. Star markers indicate the time at which each star particle forms. The shaded region encompasses the 25th to 75th percentile (average; dashed line) of the gas density for star forming gas within the inner 2\,kpc region of the galaxy. Most clump particles are initially non-star forming but end up surpassing the average density of star-forming gas in the galaxy by more than one order of magnitude.}
\label{fig:random10_nH}
\end{figure}

We can perform a more detailed analysis of clump formation compared to regular star formation by tracking the full time evolution of individual gas particles that end up forming stars in the clump. In Figure \ref{fig:random10_nH}, we take ten randomly selected particles from Clump A and track their gas densities as a function of time. We compare these gas densities to the average density for star-forming gas in the galaxy during the first $5$\,Myr of the simulation (as representative of more normal star formation before AGN winds completely disrupt the star-forming gas reservoir). Specifically, we consider star-forming gas within 2\,kpc and calculate the $[25^{\rm th}-75^{\rm th}]$ percentiles of gas density at each time. For clarity, we smooth the resulting gas density histories by applying a running average with a time window of 0.05\,Myr. At the start of the simulation, most progenitor clump particles have densities below that of the average star-forming gas in the galaxy, and often well below the threshold for star formation ($n_{\rm H,th} = 10^{3}\,{\rm cm}^{-3}$). At $\sim$1.5\,Myr after the beginning of the quasar phase, gas densities rise sharply as the clump begins to form, surpassing the average density of star-forming gas by more than one order of magnitude and resulting in the formation of stars (denoted by the star markers) at densities $n_{\rm H} \sim 10^{5}\,{\rm cm}^{-3}$, as expected from Figure \ref{fig:clusters_e1f02_nH}.

\begin{figure}
\includegraphics[width=0.5\textwidth]{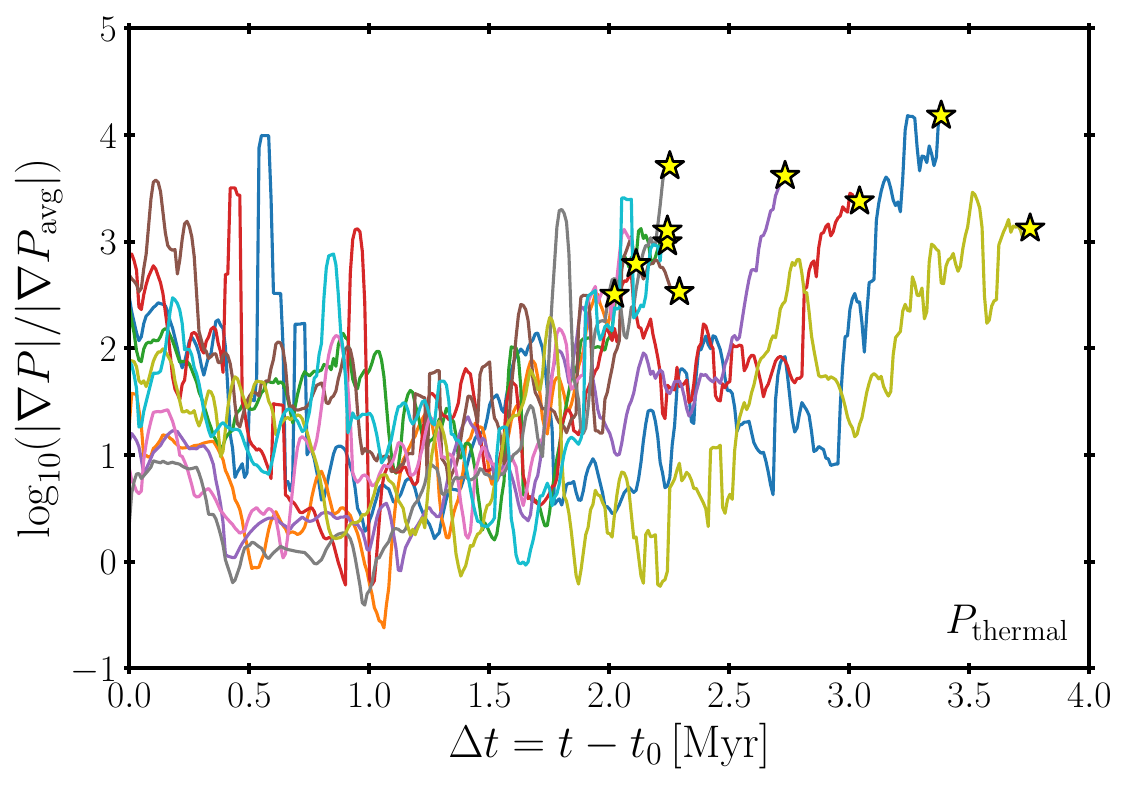}
\includegraphics[width=0.5\textwidth]{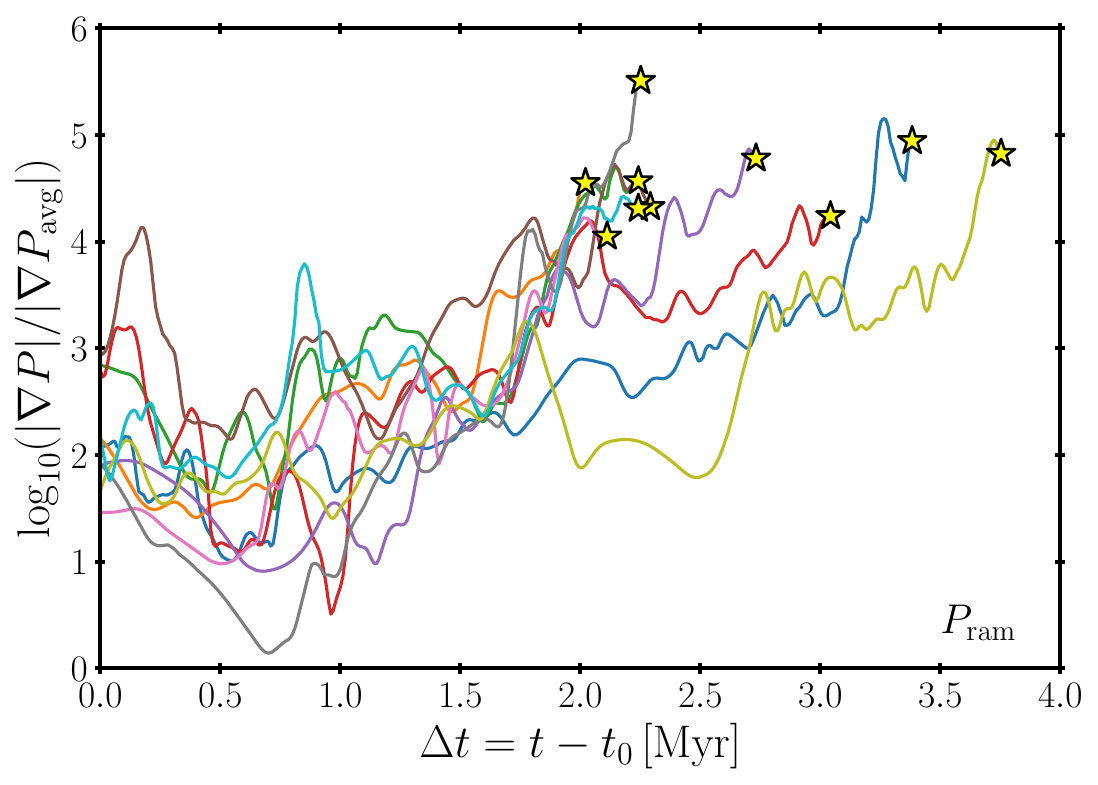}
\vspace*{-3mm}
\caption{Ratio of the magnitude of the thermal pressure gradient (top) and ram pressure gradient (bottom) on clump progenitor gas particles to that of the average pressure gradient for all star-forming gas within the central 2 kpc. Lines of different colours show the time evolution for the same Clump A star particles as in Figure \ref{fig:random10_nH}. Clump progenitor gas particles experience significantly larger pressure gradients than average star-forming gas and increasingly so as they approach their conversion into clump stars.}
\label{fig:random10_gradP} 
\end{figure}

Figures \ref{fig:clusters_e1f02_nH}-\ref{fig:random10_nH} show that stars forming in the identified clumps do so at higher densities, by over an order of magnitude, than the overall population of stars that formed under the presence of AGN winds. This qualitative difference could be attributed to different physical mechanisms or events affecting progenitor clump particles. One way to quantify the plausible impact the AGN-driven winds on clump particles is to study the difference in pressure gradients compared to the remaining star-forming gas in the galaxy. 

Figure \ref{fig:random10_gradP} shows the magnitude of the thermal pressure gradient (top) and ram pressure gradient (bottom) acting on Clump A progenitor gas particles relative to regular star-forming gas, where we follow the time evolution of the same 10 particles and apply the same 0.05\,Myr running average as in Figure \ref{fig:random10_nH}. We compute the thermal pressure as $P_{\rm thermal}=(\gamma -1)\,\rho\,U$, where $\rho$ is the gas density, $U$ is the internal energy for the gas, and $\gamma$ is the adiabatic index which we set as $5/3$. For the ram pressure, we simply take $P_{\rm ram}=\rho \abs{v}^{2}$, where $\abs{v}$ is the magnitude of the fluid velocity at the location of each gas particle. For both $P_{\rm thermal}$ and $P_{\rm ram}$, we consider all gas particles (AGN winds and pre-existing ISM gas) and compute pressure gradients in post-processing using \textsc{Meshoid}.\footnote{The \textsc{Meshoid} Python repository is available at \url{https://github.com/mikegrudic/meshoid}.}

Both panels show that Clump A progenitor particles experience up to four and five orders of magnitude larger thermal and ram pressure gradients, respectively, compared to regular star-forming gas in the galaxy. We show below that ram pressure greatly dominates over thermal pressure (\S\ref{sec:AGNdriveCF}) which, together with the increased over-pressurization leading to the final conversion of gas into stars relative to average suggests that AGN winds play a key role triggering clump formation. We can make two clear distinctions between both panels: (i) the boosting of the ram pressure is shifted up by an order of magnitude, and (ii) the evolutionary tracks of each particle for the ram and thermal pressure gradients  exhibit a similar trend up until the particle turns into a star particle. Overall both panels show similar results as in Figure \ref{fig:random10_nH}: the clump particles exhibit higher pressure gradients for both ram and thermal pressure compared to the star-forming galaxy average, with the notable feature of showing that at all times the particles are experiencing steeper pressure gradients as opposed to the star-forming gas in the galaxy on average.

\section{AGN Winds as Primary Driver of Clump Formation}\label{sec:AGNdriveCF}

Using the IDs for each particle within Clump A allows us to track them back in time to follow their evolution from a gas element up until they form into a star particle (as shown in \S\ref{subsec:beforeSF}), but we can also track the same particles between simulations of varying feedback strengths as well as in our \nf~simulation. This provides us with the opportunity to compare various properties between the two simulation runs and identify differences that can point to the plausible effect of AGN-driven winds on the formation of the clump. By tracking the same particles between both the \mf~and \nf~simulations we can gain further insight into the reasons behind the absence of the clump in the \nf~simulation.

\begin{figure}
\includegraphics[width = \columnwidth]{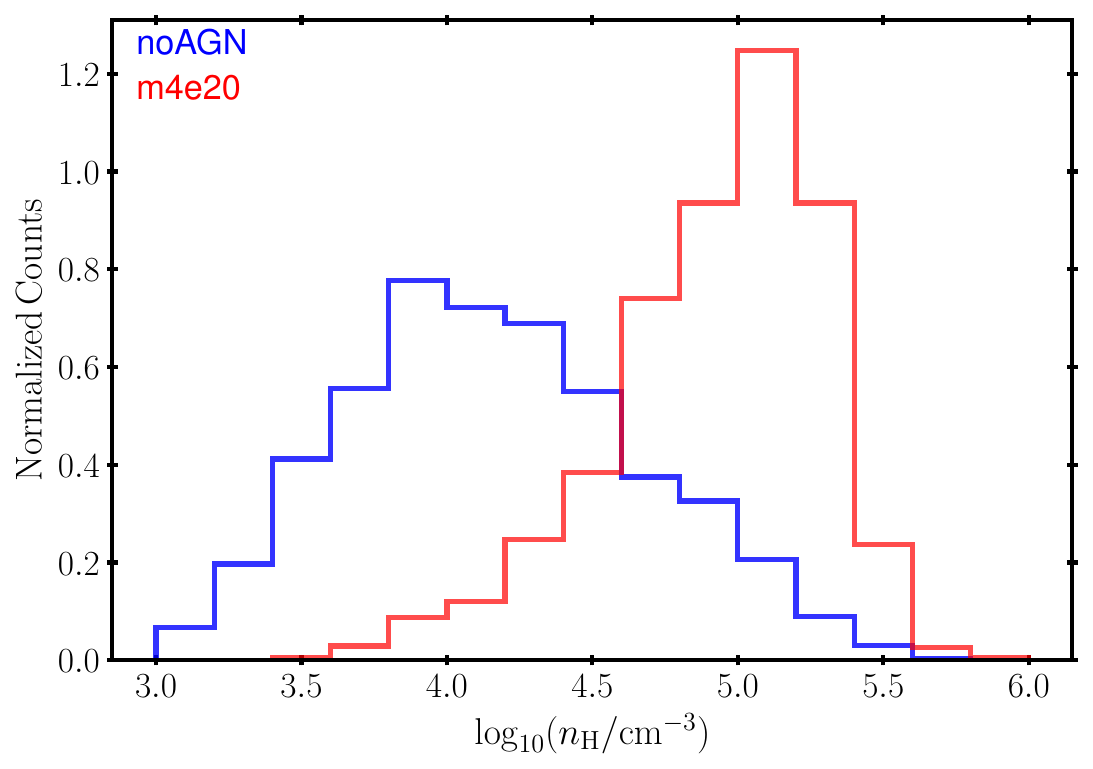}
\vspace*{-5mm}
\caption{Normalized probability distribution of the gas density ($n_{\rm H}$) at which Clump A star particles form in the \mf~AGN wind simulation (red; as in Figure \ref{fig:clusters_e1f02_nH}) compared to the density at which the same gas particles (identified by ID) turn into star particles in the \nf~simulation (blue). Most Clump A progenitor gas particles also turns into star particles in the \nf~simulation but at significantly lower densities in the absence of AGN winds.}
\label{fig:NFvsF_histogram} 
\end{figure}

\begin{figure}
\includegraphics[width = \columnwidth]{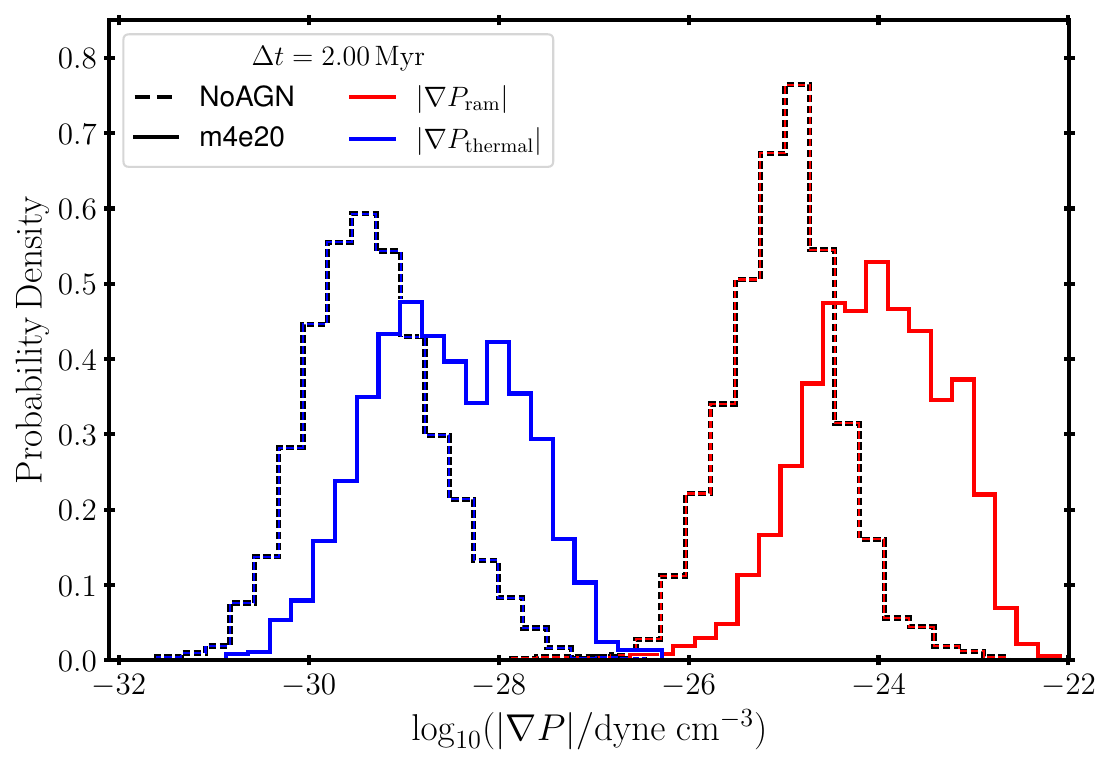}
\vspace*{-4mm}
\caption{Normalized probability density distribution of the pressure gradient norm for ram pressure (red) and thermal pressure (blue) at $\Delta t=2$\,Myr for gas particles that end up forming Clump A stars in the \mf~simulation (solid) compared to the pressure gradient distributions for the same particles tracked in the \nf~simulation (dashed).
Ram pressure gradients are always larger than thermal pressure gradients for both simulations. However, Clump A particles form subject to larger pressure gradients in the presence of strong AGN winds compared to the same particles in the \nf~simulation.}
\label{fig:NFvsF_gradPs}
\end{figure}

\begin{figure*}
\includegraphics[width = \textwidth]{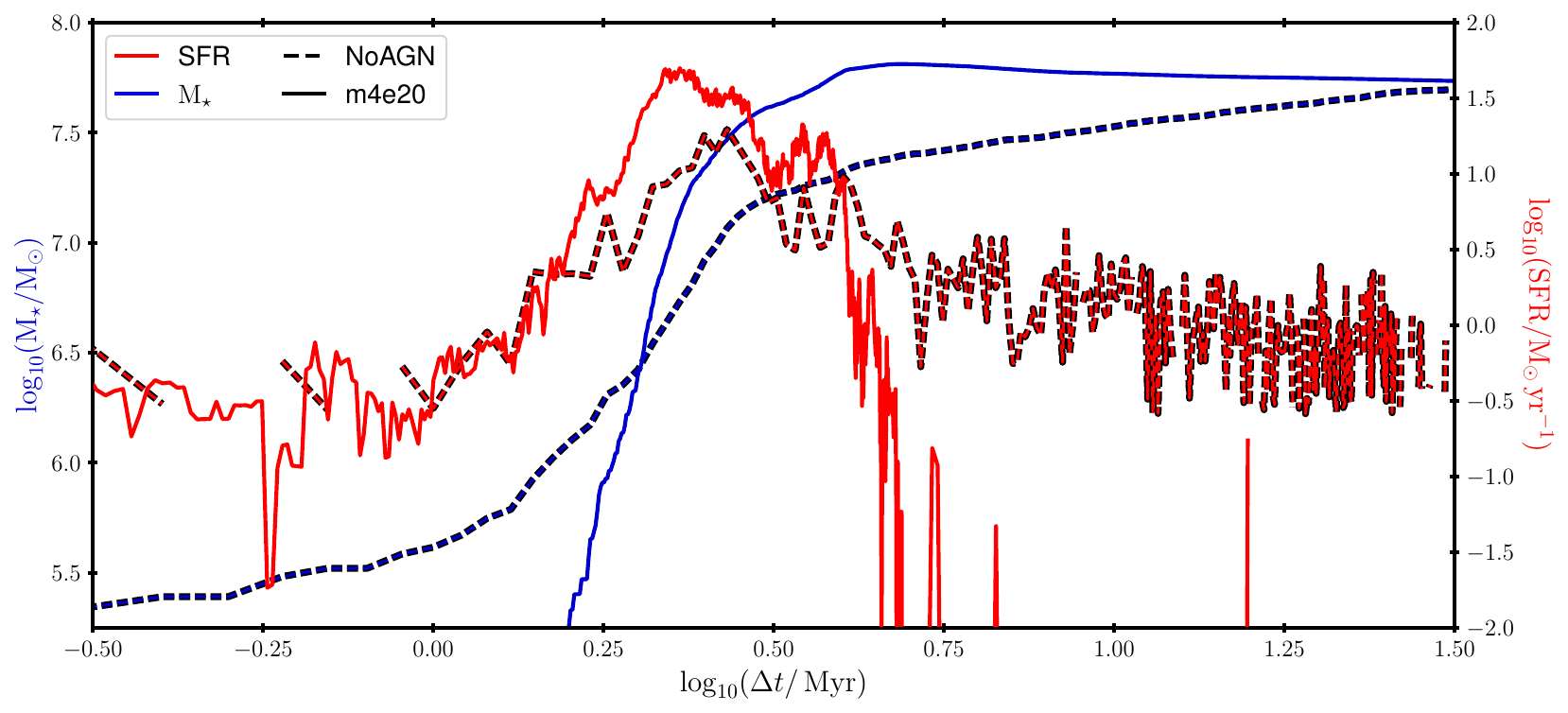}
\vspace*{-4mm}
\caption{Stellar mass growth of Clump A (blue; left axis) and corresponding SFR of progenitor gas particles (red; right axis) as a function of time in simulation \mf~(solid) compared to the build up of stellar mass and SFR from the same gas particles identified by ID in the \nf~simulation (dashed). Most stars in Clump A form in a very short burst reaching ${\rm SFR}\sim 50\,\Msunyr$ at $\Delta t\sim 2\,{\rm Myr}$ since the start of the quasar wind phase, while the same gas elements form into stars over a much longer period of time in the \nf~simulation.}
\label{fig:SFRvstime} 
\end{figure*}

\subsection{Impact of AGN winds on density and pressure gradients}\label{subsec:AGNonprops}

Due to the high efficiency of star formation in the \nf~simulation, we find that a majority of the progenitor Clump A particles that turned into star particles in the \mf~simulation also do so in the absence of AGN winds. 
Figure \ref{fig:NFvsF_histogram} investigates the formation of these stars in the \nf~case compared to the \mf~simulation by showing the normalized probability distribution of gas density $n_{\rm H}$ just prior to forming into stars in each simulation.
Although the \nf~simulation forms $\sim$95\,\% of the stars from the same gas progenitor particles of Clump A, they form at significantly lower densities (more than an order of magnitude) compared to the strong AGN wind case. The $n_{\rm H}$ distribution for progenitor clump particles in the \nf~simulation is broadly consistent with the overall population of stars (Figure \ref{fig:35MyrallSIMS}), and we show below (\S\ref{subsec:ClumpGrowth}) that they do not form a clump or coherent stellar structure in the absence of AGN winds.

Figure \ref{fig:NFvsF_gradPs} shows the normalized probability density distribution of the magnitude of the ram pressure (red) and thermal pressure (blue) gradients for Clump A progenitor gas particles at $\Delta t= 2$\,Myr in simulation \mf~(solid lines) compared to the same gas particles in the \nf~simulation (dashed lines). We choose $\Delta t=2$\,Myr as a representative time at which Clump A is rapidly forming stars while it is still very gas rich, allowing us to perform a statistical comparison of gas particle properties in simulations with and without AGN winds (our results are not sensitive to this specific choice).
We find that the ram pressure gradients are always stronger than the thermal pressure gradients by more than four orders of magnitude in both simulations. Importantly, the pressure gradient distributions are very different in the \mf~and \nf~simulations, with the same clump progenitor gas particles experiencing much stronger pressure gradients in the presence of AGN winds. At this time ($\Delta t= 2$\,Myr), Clump A is rapidly forming stars in simulation \mf~while the same gas elements fuel significantly lower SFR in the \nf~simulation (shown explicitly in \S\ref{subsec:ClumpGrowth}), pointing to pressurization by strong AGN winds as a key driver of the formation of the identified ultra dense stellar clumps. 

\subsection{Clump Growth and Size Evolution}\label{subsec:ClumpGrowth}
Figure \ref{fig:SFRvstime} shows the SFR (red) and stellar mass growth (blue) of Clump A in simulation \mf~as a function of time (solid lines), tracking the clump progenitor particles, compared to the SFR and build up of stellar mass of the same gas particles identified by ID in the \nf~simulation (dashed lines).
We see that the progenitor clump particles actually begin to form stars earlier in the \nf~simulation while AGN winds appear to suppress their SFR in the first $\sim$1\,Myr. However, as the AGN winds compress the progenitor gas particles, a short burst of star formation reaching ${\rm SFR}\sim 50\,\Msunyr$ at $\Delta t\sim 2$\,Myr forms most of the stellar mass of Clump A in less than $\sim$4\,Myr. Meanwhile, the same gas particles in the \nf~simulation continue to form stars at a slower rate during $\sim$35\,Myr, providing further indication that strong AGN winds are required for such an extreme clump formation event to occur. By the end of the simulation ($\Delta t = 35\,{\rm Myr}$), we see that both the \mf~and \nf~simulations reach roughly similar stellar mass out of the same parent gas particles but under rather different conditions.

\begin{figure*}
\includegraphics[width = \textwidth]{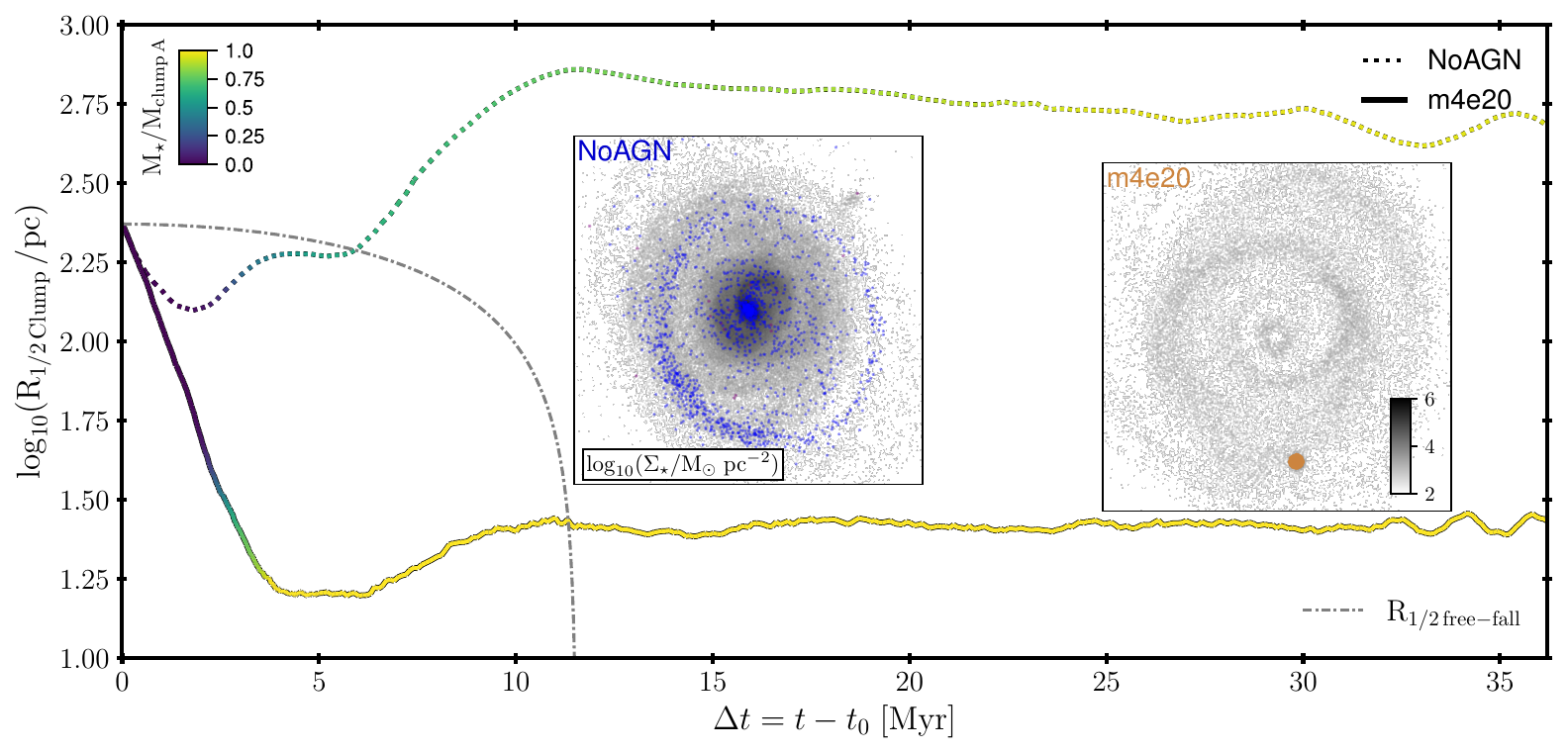}
\vspace*{-4mm}
\caption{Radius enclosing half of the mass of Clump A progenitor particles as a function of time in simulation \mf~ (solid) compared to the same particles identified by ID in the \nf~simulation (dashed). The colour scale for each line indicates the fraction of Clump A progenitor particles that have turned into stars as a function of time. The dash-dotted line (gray) shows the half mass radius as a function of time for an idealized spherical gas cloud with uniform density and the same mass and initial half mass radius as Clump A, collapsing under its own gravity and neglecting all other forces. Inset panels show the face-on stellar mass surface density distribution superimposed with the location of Clump A star particles at $\Delta t= 35\,{\rm Myr}$ for simulation \mf~(right inset) compared to the same particles identified in the \nf~simulation (left inset). Compression by AGN winds makes Clump A form significantly faster than the dynamical time to collapse under its own gravity.}
\label{fig:separation} 
\end{figure*}

Figure \ref{fig:separation} further explores the conditions that are driving the formation of Clump A under the presence of strong AGN winds, while the clump is absent in the \nf~simulation. For all Clump A progenitor particles (either gas or stars already formed), we compute the radius containing half of their total mass relative to their center of mass at each time. This effective clump radius, ${\rm R}_{\rm 1/2\,Clump}$, is thus a measure of how compact or spread the progenitor gas cloud is as the clump forms. Figure \ref{fig:separation} shows the effective radius of Clump A as a function of time (solid line) compared to ${\rm R}_{\rm 1/2\,Clump}$ calculated for the same set of progenitor particles identified in the \nf~simulation (dotted line), with the fraction of mass in the form of stars encoded by the colour scale in each case. For reference, we also plot the half mass radius ${\rm R}_{\rm 1/2\,free-fall}$ versus time for an idealized spherical gas cloud with the same mass and initial size as Clump A, uniform density, initially at rest, and collapsing under its own gravity neglecting all other forces (gray dash-dotted line). Following the derivation of the free-fall time, applying Newton's second law to the equation of motion for a test particle at the edge of a cloud we instead solve for the radius. We numerically solve the second-order differential equation $\ddot{{\rm R}}_{\rm 1/2\,free-fall}(\Delta t) = -0.5GM_{\rm Clump}/{\rm R}_{\rm 1/2\,free-fall}$, with initial conditions (i) ${\rm R}_{\rm 1/2\,free-fall}(\Delta t = 0) = {\rm R}_{\rm 1/2\,Clump}(\Delta t = 0)$ and (ii) $\dot{{\rm R}}_{\rm 1/2\,free-fall}(\Delta t = 0) = 0$. This free-fall time thus represents the shortest amount of time that Clump A would require to form in the absence of external forces and neglecting support from thermal pressure, turbulence, internal rotation, or shear forces. 

Under the presence of strong AGN winds, the progenitor gas cloud is quickly compressed from ${\rm R}_{\rm 1/2\,Clump} \sim 250\,{\rm pc}$ down to $<$20\,pc in less than 4\,Myr, while most of the clump gas is converted into stars once ${\rm R}_{\rm 1/2\,Clump} < 30\,{\rm pc}$ at $\Delta t > 2\,{\rm Myr}$.  In contrast, an idealized gas cloud with similar mass and size collapsing under its own weight would require $>$10\,Myr to form a dense clump even neglecting any forces that could provide support against gravitational collapse. Compression by strong AGN winds therefore appears as a key ingredient for the fast formation of ultra dense stellar clumps. In the absence of AGN winds, the progenitor gas cloud collapses more slowly and less coherently into different structures, with a significant fraction of gas forming stars in the nuclear region and an off-center, lower-density clump that ends up disrupted by tidal forces as illustrated in the inset panel.

\section{Discussion} \label{sec:discussion}

In \citet{Mercedes-Feliz2023}, we showed that AGN winds powered by a rapidly accreting central BH in FIRE simulations of a massive star-forming galaxy at the peak of activity can have global negative effects (suppressing star formation) for a range of assumed kinetic feedback efficiencies.  We also identified several different signatures of local positive AGN feedback, including higher local star formation efficiency in compressed gas along the central cavity and the presence of outflowing material with ongoing star formation, but we showed that in all cases the negative AGN feedback effects always dominate, suppressing more than triggering star formation. The detailed analysis presented here further supports these conclusions for the case of very strong quasar winds, in qualitative agreement with the overall negative effects of previous AGN feedback implementations in galaxy formation simulations \citep{Choi2015,Schaye2015,Hirschmann2016,Angles-Alcazar2017a,Tremmel2017,Weinberger2017,Dave2019,Habouzit2021,Habouzit2022,Byrne2023formation,Wellons2023} but in contrast with some analytic models and idealized simulations suggesting that AGN feedback could have net positive effects and even drive strong starbursts \citep{Gaibler2012,ishibashi2012,Zubovas2012,silk2013,zubovas2013,nayakshin2014,bieri2015,bieri2016,zubovas2017}.

The ultra-dense stellar clumps analysed here, with $\sim$10$^7\,\Msun$ of stars packed into $\sim$20\,pc, represent the most extreme examples of local positive AGN feedback identified in our simulations.  By carefully tracking back in time their formation and identifying the same progenitor gas cloud in an identical simulation without AGN winds, we have shown explicitly that the rapid compression of gas by AGN winds is indeed driving the formation of these extreme stellar clumps.  Intriguingly, only the most powerful quasar winds implemented here appear to be capable of forming such clumps, corresponding to kinetic outflows with energy injection rate $\dot{E}_{\rm w} > 10^{46}$\,erg\,s$^{-1}$ produced by an Eddington-limited BH with $M_{\rm BH} = 10^9\,\Msun$ and kinetic efficiency $\epsilon_{\rm k} > 0.1$. Luminous red quasars exhibit bolometric luminosities reaching $10^{47-48}\,{\rm erg}\,{\rm s}^{-1}$ \citep[e.g.,][]{Goulding2018}, with inferred outflow energies spanning $3-50\%$ of the quasar luminosity \citep[e.g.,][]{Perrotta2019,Heckman2023} which are thus roughly consistent with the strongest quasar winds modeled here. Nonetheless, collimated winds or jets, as opposed to isotropic winds, could have a similar effect compressing gas clumps at lower net energy output.

In our simulations, we inject quasar winds at the time that the host galaxy is undergoing its strongest starburst phase and gravitational torques from multi-scale stellar non-axisymmetries can indeed drive quasar-like gas inflow rates down to sub-pc scales \citep{Angles-Alcazar2021}. The host galaxy satisfies observational constraints at higher redshift, including BH-galaxy, stellar mass-halo mass, and galaxy mass-size relations \citep{Angles-Alcazar2017c,Feldmann2016,Feldmann2017,Wellons2020}, but the $z\sim2.3$ starburst leads to the formation of an overcompact and overdense stellar component in the absence of AGN winds since stellar feedback is no longer able to regulate star formation \citep{Wellons2020,Parsotan2021}.
Our simulations injecting strong quasar winds are able to  quench star formation in the host galaxy in $\sim$20\,Myr \citep{Cochrane2023,Mercedes-Feliz2023,Angles-Alcazar2023}, maintaining the galaxy in agreement with observed stellar surface densities and the mass-size relation \citep{Cochrane2023} while leaving one or two off-center ultra-dense stellar clumps as a direct signature of positive AGN feedback during the global quenching process. Our simulations are thus consistent with available observational constraints and support the proposed AGN-driven dense star cluster formation scenario. 

Observational and theoretical works indicate different pathways for the formation of star clusters in galaxies, including gravitational instabilities, turbulent fragmentation, merging of smaller clumps, and hierarchical assembly \citep{Rieder2013,Inoue2016,Kim2018,Webb2019,Li2019,Adamo2020,Phipps2020,Garcia-Bernete2021,Grudic2021,Faisst2022,Han2022,Li2022,Reina-Campos2022,Larson2023,Sameie2023}. Massive, gas-rich galaxies at cosmic noon ($z\sim2$) often show signs of extended rotating discs along with giant star-forming clumps that can reach masses $\sim$10$^{7-9}\,\Msun$ and sizes $\sim$100--1000\,pc \citep{Elmegreen2009,Wuyts2012,Guo2015,Huertas-Company2020}. Hydrodynamic simulations of gas-rich discs generally show that such clumps can either originate from infalling satellites or form via disc instabilities with a mass near or below the characteristic Toomre mass  \citep{Genel2012,Hopkins2012_clumps,Angles-Alcazar2014,Moody2014,Mandelker2017,Oklopcic2017,Ma2018,Dekel2022}, where the fate of these clumps (whether they quickly disrupt or slowly sink to form the central bulge) depends on resolution and stellar feedback implementation \citep{Genel2012,Hopkins2012_clumps,Oklopcic2017,Ceverino2023}.
Galaxy mergers are another proposed pathway to form stellar clumps due to the high-pressure, gas-rich environments produced
\citep{Renaud2008a,Renaud2008b,Teyssier2010,Herrera2011,Renaud2015,Kim2018,Moreno2019,vanDonkelaar2023b}.
Our simulations without AGN winds also produce massive, star-forming clumps through turbulent fragmentation and gravitational instability, but these are typically short-lived and quickly disrupted by radiative feedback, in agreement with previous FIRE simulations \citep{Oklopcic2017,Ma2018,Ma2020}.

The AGN feedback-triggered stellar clump formation scenario identified in our simulations is, however, clearly distinct from the traditional star-forming clumps formed in gas-rich galaxies via gravitational instability or mergers. With stellar mass ${\rm M}_{\star}\sim 10^{7}\,\Msun$ and surface density $\Sigma_{\star} \sim 10^{4}\,\Msunpc$, the stellar clumps analyzed here have properties similar to observed nuclear star clusters, super star clusters, and ultra compact dwarfs in the low-$z$ Universe \citep{Bastian2013,Norris2014,Grudic2019} as well as some of the densest clumps observed in lensed systems at $z \sim 2-8$ \citep{Bouwens2021,Mestric2022}. 
A crucial aspect of this positive AGN feedback clump formation scenario is that AGN winds compress the progenitor gas cloud to much higher densities and much faster that could happen otherwise, with the gas cloud shrinking by a factor $\sim$1000 in volume and converting most of its mass into stars in only $\sim$2\,Myr.
The immediate impact of radiative feedback from massive stars is the dominant cloud dispersion mechanism before the first SNe go off, but the clump mass surface density is high enough to prevent disruption. Theoretical and observational studies have shown that the star formation efficiency (SFE) in collapsing clouds scales with the total mass surface density \citep[$\Sigma_{\rm tot}$;][]{Hopkins2012,Grudic2018,Kim2018_radiative,Wong2019}, with clouds reaching $\Sigma_{\rm tot} > \Sigma_{\rm crit} \approx 1000\,\Msunpc$ expected to turn most of their gas into stars as gravity overcomes the total momentum input from stellar feedback \citep{Grudic2018,Grudic2020} and the fraction of stars formed in a gravitationally bound cluster also quickly rising at high SFE \citep{Grudic2021}.
In our simulations, strong AGN winds trigger the formation of gas clumps with densities much higher than $\Sigma_{\rm crit}$, resulting in the formation of massive, gravitationally bound stellar clumps with SFE\,$\sim$\,1.

Previous simulations have argued that other physical mechanisms besides self-gravity are helping giant molecular clouds collapse and form stars in starburst galaxies \citep{Ma2020,He2023}, and violent mergers of proto-galaxies at high-redshift have been proposed as a viable mechanism to quickly form globular clusters before stellar feedback can regulate star formation in the densest gas clouds \citep{Kim2018}.
We have shown that our extreme stellar clumps form on a timescale significantly shorter than the initial free-fall time of the progenitor gas cloud, demonstrating that gravity alone cannot form these objects and that ram pressure gradients provided by strong AGN winds are a crucial ingredient.  In fact, the same progenitor gas cloud in the absence of AGN winds forms stars at significantly lower densities and over a much longer period of time ($\sim 35\,{\rm Myr}$), with the resulting stellar structure quickly losing spatial coherency owing to tidal disruption by the host galaxy. 
In contrast, the ultra-dense stellar clump formed by positive AGN feedback remains gravitationally bound for $\Delta t= 70\,{\rm Myr}$, completing a few orbits around the center of the galaxy without signs of tidal disruption. 
Positive feedback by strong AGN winds may thus represent a plausible formation scenario for globular clusters. 

Our results complement and extend previous work by \citet{Ma2020} on the formation of bound star clusters in a sample of high-resolution cosmological zoom-in simulations of $z \geq 5$ galaxies from the FIRE project. They identified gravitationally bound star clusters that form in high-pressure clouds under the influence of stellar feedback. Notably, they found that stars in clusters tend to form in gas that is one order of magnitude denser than the typical gas density at which normal star formation occurs in the host galaxy. These high-density clouds are compressed by stellar feedback-driven winds and collisions of smaller clouds in highly turbulent environments, with the cloud-scale star formation efficiency approaching unity owing to the fast formation relative to the time for internal stellar feedback to react and stop star formation. 
While we focus on a more massive host galaxy at lower redshift and undergoing a luminous quasar phase, many of our findings mirror the results of \citet{Ma2020} for the case of positive AGN feedback instead of stellar feedback-triggering of star formation.  The stellar clumps analyzed here are nonetheless representative of significantly more extreme and rarer conditions, where strong quasar winds provide just the right amount of ram pressure on a gas cloud with the optimal geometry and timing to quickly make it collapse to very high density while the remaining ISM gas content is evacuated from the host galaxy.

High gas mass resolution and adaptive gravitational softenings, explicit treatments of star formation in self-gravitating molecular gas and local stellar feedback \citep{Hopkins2018}, and hyper-refined AGN winds self-consistently capturing the geometry-dependent wind-ISM interaction \citep{Torrey2020,Angles-Alcazar2023} are all crucial ingredients to model the formation of these ultra-dense stellar clumps.
Lower resolution simulations and/or relying on pressurized ISM models where star formation occurs at much lower average densities \citep[e.g.,][]{Pillepich2018,Schaye2015,Dave2019} are thus not expected to resolve the formation of dense stellar clumps even in the presence of strong AGN winds.
With stellar gravitational softening $\epsilon_{\star}=7\,{\rm pc}$, our simulations predict stellar clump half-mass radii ${\rm R}_{\rm 1/2\, \rm Clump} \sim 3 \times \epsilon_{\star}$, suggesting that they could reach even higher densities in higher resolution simulations.  
The details of the star formation prescription \citep[e.g.,][]{Hopkins2018,Nobels2023}
may also impact the detailed properties of ultra dense stellar clumps, but they nonetheless appear to be the strongest manifestation of local positive AGN feedback in massive star-forming galaxies at their peak of activity.

Overall, our results support the dual role of AGN feedback in galaxies, which can trigger star formation locally while globally suppressing galaxy growth, and identify the conditions that can lead to the formation of ultra-dense stellar clumps and possibly globular clusters driven by powerful quasar winds. Future work should explore the AGN wind-ISM interaction and the dual role of AGN feedback for a broader range of host galaxy properties and redshifts \citep{Wellons2023,Byrne2023formation} in cosmological hyper-refinement simulations with highly resolved multi-phase ISM \citep{Angles-Alcazar2021,Hopkins2023_superzoom1}, and investigate the frequency, lifetime, and observability of ultra-dense stellar clumps driven by local positive AGN feedback.

\section{Summary and Conclusions} \label{sec:summary}
We have presented a detailed analysis of ultra-dense stellar clumps identified in a set of high-resolution cosmological zoom-in simulations of a massive galaxy near the peak of star formation activity ($M_{\rm halo} \sim 10^{12.5}\,{\rm M}_{\odot}$ at $z\sim2$) undergoing a strong quasar wind phase. The goal of this study is to investigate the implications of AGN feedback on the formation of these stellar structures and to further investigate the plausible positive versus negative effects of AGN feedback during a luminous quasar phase \citep{Mercedes-Feliz2023}. Our simulations include local stellar feedback and resolved multi-phase ISM physics from the FIRE-2 project \citep{Hopkins2018}, as well as hyper-refined AGN-driven winds which simultaneously capture their propagation and impact from the inner few pc to CGM scales \citep{Angles-Alcazar2023}. Our main results can be summarized as follows:
\begin{enumerate}[wide, labelwidth=!,itemindent=!]
    \item Only simulation variants with very strong AGN winds (mechanical energy injection rate $\dot{E}_{\rm w} > 10^{46}$\,erg\,s$^{-1}$) lead to the formation of ultra-dense, off-center clumps with stellar mass ${\rm M}_{\star}\sim 10^{7}\,\Msun$, effective radius ${\rm R}_{\rm 1/2\, \rm Clump}\sim 20\,{\rm pc}$, and surface density $\Sigma_{\star} \sim 10^{4}\,\Msunpc$. Collimated (as opposed to isotropic) outflows could have a similar effect at lower net energy output.
    \item Star particles that formed within the clumps do so at significantly higher gas density than the overall population of stars formed during the same time, reaching $n_{\rm H}\sim 10^{5}\,{\rm cm}^{-3}$ or roughly two orders of magnitude above the density threshold for star formation ($n_{\rm H, th} = 10^{3}\,{\rm cm}^{-3}$).
    \item Progenitor clump particles are typically below the star formation threshold but increase their density rapidly owing to ram pressure gradients orders of magnitude larger than the pressure gradients experienced by regular star-forming gas in the galaxy.
    \item Tracking the same clump progenitor particles in the \nf~simulation, we demonstrate that most of that gas also forms stars but at significantly lower densities and experiencing much weaker pressure gradients compared to the same gas cloud in the presence of strong AGN winds.
    \item Rapid compression of gas by AGN winds drives a strong burst of star formation reaching ${\rm SFR} \sim 50\,\Msunyr$ and $\Sigma_{\rm SFR} \sim  10^{4}\,\Msunyrkpc$ which converts most of the progenitor gas cloud into gravitationally bound stars in $\sim$2\,Myr, with stellar feedback unable to regulate star formation. In contrast, the same gas cloud in the absence of AGN winds forms stars over a much longer period of time ($\sim 35\,{\rm Myr}$) and losing spatial coherency.
    \item The rate at which the progenitor gas cloud collapses to form the stellar clump (${\rm R}_{\rm 1/2\, \rm Clump} \approx 250 \rightarrow 20\, {\rm pc}$ in $\sim 2\,{\rm Myr}$) is much faster than the free-fall time under its own gravity even neglecting internal pressure support, turbulence, or shear forces, further emphasizing the need for strong AGN winds to enable the formation of these extreme stellar clumps.
\end{enumerate}

Our results suggest that young, ultra-dense stellar clumps in recently quenched galaxies could be a unique signature of local positive AGN feedback acting alongside strong negative feedback by quasar winds,
providing a plausible formation scenario for globular clusters.

\section*{Acknowledgements}
We thank the anonymous referee for constructive comments that helped improve the paper. The simulations were run on Flatiron Institute’s research computing facilities (Gordon-Simons, Popeye, and Iron compute clusters), supported by the Simons Foundation. We thank the Scientific Computing Core group at the Flatiron Institute for outstanding support.
Additional numerical calculations were run on the Caltech compute cluster “Wheeler,” allocations FTA-Hopkins supported by the NSF and TACC, and NASA HEC SMD-16-7592, and XSEDE allocation TG-AST160048 supported by NSF grant ACI-1053575.
JMF was supported in part by a NASA CT Space Grant Graduate Fellowship.
DAA acknowledges support by NSF grants AST-2009687 and AST-2108944, CXO grant TM2-23006X, JWST grant GO-01712.009-A, Simons Foundation Award CCA-1018464, and Cottrell Scholar Award CS-CSA-2023-028 by the Research Corporation for Science Advancement.
JM is funded by the Hirsch Foundation.
CAFG was supported by NSF through grants AST-2108230, AST-2307327, and CAREER award AST-1652522; by NASA through grants 17-ATP17-0067 and 21-ATP21-0036; by STScI through grant HST-GO-16730.016-A; and by CXO through grant TM2-23005X.
Support for PFH was provided by NSF Research Grants 1911233, 20009234, 2108318, NSF CAREER grant 1455342, NASA grants 80NSSC18K0562, HST-AR-15800.

\section*{Data Availability}
The data supporting the plots within this article are available on reasonable request to the corresponding author. FIRE-2 simulations are publicly available \citep{Wetzel2023} at \url{http://flathub.flatironinstitute.org/fire}. Additional FIRE simulation data, including initial conditions and derived data products, are available at \url{https://fire.northwestern.edu/data/}. A public version of the GIZMO code is available at \url{http://www.tapir.caltech.edu/~phopkins/Site/GIZMO.html}.



\bibliographystyle{mnras}
\bibliography{main} 








\bsp	
\label{lastpage}
\end{document}